\documentclass[12pt,dvips]{article}
\usepackage{rotating}
\usepackage{axodraw}
\usepackage{epsfig}
\usepackage{amssymb}
\usepackage{color}
\newcommand{\imag}{\Im {\rm m}}
\newcommand{\real}{\Re {\rm e}}

\def\lsim{\mathrel{\raise.3ex\hbox{$<$\kern-.75em\lower1ex\hbox{$\sim$}}}}
\def\gsim{\mathrel{\raise.3ex\hbox{$>$\kern-.75em\lower1ex\hbox{$\sim$}}}}
\newcommand{\neu}{\tilde{\chi}^0}
\begin{document}

\renewcommand{\thefootnote}{\fnsymbol{footnote}}

\begin{titlepage}

\begin{flushright}
DESY 06-016 \\[-0.1cm]
KIAS--P06005 \\[-0.1cm]
hep-ph/0602131\\[5mm]
%{\it SYC on \today}
\end{flushright}

\vspace{0.5cm}

\begin{center}
  {\Large \bf Neutralino Production and Decay at an \boldmath{$e^+e^-$} Linear
   Collider with Transversely Polarized Beams}\\[1.cm]
   {\large S.Y. Choi$^{1}$, M. Drees$^2$, and J. Song$^3$}
\end{center}

\vskip 0.5cm

{\small
\begin{center}

$^1$ {\it Deutsches Elektronen--Synchrotron DESY, 22603 Hamburg, Germany}\\
          and\\
{\it Department of Physics and RIPC, Chonbuk National University, Jeonju
  561-756, Korea}\footnote{Permanent Address}\\[2mm]
$^2$ {\it KIAS, School of Physics, Seoul 130--012, Korea} \\ and \\
{\it Physikalisches Institut, Universit\"at Bonn, Nussallee 12, D53115 Bonn,
          Germany}\footnote{Permanent Address} \\[2mm]
$^3$ {\it Department of Physics, Konkuk University, Seoul 143--701,
          Korea}
\end{center}
}

\vspace{1.cm}

\begin{abstract}
\noindent Once supersymmetric neutralinos $\tilde{\chi}^0$ are
produced copiously at $e^+e^-$ linear colliders, their
characteristics can be measured with high precision. In particular,
the fundamental parameters in the gaugino/higgsino sector of the
minimal supersymmetric extension of the standard model (MSSM) can be
analyzed. Here we focus on the determination of possible CP--odd
phases of these parameters.  To that end, we exploit the
electron/positron beam polarization, including transverse
polarization, as well as the spin/angular correlations of the
neutralino production $e^+ e^- \to \tilde{\chi}^0_i
\tilde{\chi}^0_j$ and subsequent 2--body decays $\tilde{\chi}^0_i
\to \tilde{\chi}^0_k h, \tilde{\chi}^0_k Z, \tilde \ell^\pm_R \ell
^\mp$, using (partly) optimized CP--odd observables. If no
final--state polarizations are measured, the $Z$ and $h$ modes are
independent of the $\tilde{\chi}^0_i$ polarization, but CP--odd
observables constructed from the leptonic decay mode can help in
reconstructing the neutralino sector of the CP--noninvariant MSSM.
In this situation, transverse beam polarization does not seem
to be particularly useful in probing explicit CP violation in the
neutralino sector of the MSSM. This can most easily be accomplished
using longitudinal beam polarization.

\end{abstract}

\vskip 0.5cm

\end{titlepage}

\renewcommand{\thefootnote}{\fnsymbol{footnote}}

%========================
\section{Introduction}
\setcounter{footnote}{0}
\label{sec:introduction}
%========================

In the minimal supersymmetric standard model (MSSM) \cite{book}, the spin-1/2
partners of the neutral gauge bosons, $\tilde{B}$ and $\widetilde{W}_3$, and
of the neutral Higgs bosons, $\tilde H_1^0$ and $\tilde H_2^0$, mix to form
the neutralino mass eigenstates $\chi_i^0$ ($i$=1,2,3,4). The corresponding
mass matrix in the $(\tilde{B},\widetilde{W}_3,\tilde{H}^0_1,\tilde{H}^0_2)$
basis
\begin{eqnarray}
{\cal M}=\left(\begin{array}{cccc}
  M_1       &      0          &  -m_Z c_\beta s_W  & m_Z s_\beta s_W \\[2mm]
   0        &     M_2         &   m_Z c_\beta c_W  & -m_Z s_\beta c_W\\[2mm]
-m_Z c_\beta s_W & m_Z c_\beta c_W &       0       &     -\mu        \\[2mm]
 m_Z s_\beta s_W &-m_Z s_\beta c_W &     -\mu      &       0
                  \end{array}\right)\
\label{eq:massmatrix}
\end{eqnarray}
contains several fundamental supersymmetry parameters: the U(1) and SU(2)
gaugino masses $M_1$ and $M_2$, the higgsino mass parameter $\mu$, and the
ratio $\tan\beta=v_2/v_1$ of the vacuum expectation values of the two neutral
Higgs fields. Here, $s_\beta =\sin\beta$, $c_\beta=\cos\beta$ and $s_W,c_W$
are the sine and cosine of the electroweak mixing angle $\theta_W$.

In CP--noninvariant theories, the mass parameters $M_{1,2}$ and $\mu$ are
complex.  By re-parameterizing the fields, $M_2$ can be taken
real and positive without loss of generality.
Two remaining
non--trivial phases are attributed to $M_1$ and $\mu$:
\begin{eqnarray}
M_1=|M_1|\,\,{\rm e}^{i\Phi_1}\ \qquad {\rm and} \qquad
\mu=|\mu|\,\,{\rm e}^{i\Phi_\mu} \qquad (0\leq \Phi_1,\Phi_\mu< 2\pi) \, .
\end{eqnarray}
The existence of CP--violating phases in supersymmetric theories induces, in
general, electric dipole moments (EDM) \cite{edm}. The current experimental
bounds on the EDM's constrain the parameter space including
many parameters outside the neutralino/chargino sector \cite{edm1}.  Detailed
analyses of the electron EDM show \cite{edm1,Choi:2004rf} that the phase
$\Phi_\mu$ must be quite small, unless selectrons are very
heavy.\footnote{Large values of $\Phi_\mu$ can also be tolerated for moderate
  selectron masses if $\tan\beta$ is close to 1.  However, this possibility is
  essentially excluded by Higgs boson searches at LEP.} In contrast, large
values of $\Phi_1$ are allowed even for rather small selectron masses.  The
CP--violating phase $\Phi_1$ can therefore play a significant role in the
production and decay of neutralinos, which is most easily investigated at
(linear) $e^+e^-$ colliders \cite{others, CSS,cdgs,Choi:2004rf,newbartl}.

Neutralinos are produced in $e^+e^-$ collisions, either in diagonal or mixed
pairs \cite{oldino}. If the collider energy is high enough to produce all four
neutralino states, the underlying SUSY parameters $\{|M_1|, \Phi_1,
M_2,\,|\mu|, \Phi_{\mu}; \tan\beta\}$ can be extracted from the masses
$m_{\tilde{\chi}^0_i}$ ($i$=1,2,3,4) and the cross sections \cite{rec,
  Choi:2001ww}. At the first stage of operations of a linear $e^+e^-$
collider, however, only the lighter neutralinos may be accessible.
If $\tilde \chi_1^0 \tilde \chi_2^0$ is the only visible neutralino
pair that is accessible, measuring their masses and (polarized)
production cross sections may not suffice to determine the
parameters of the neutralino mass matrix completely; the detailed
analysis of $\tilde \chi_2^0$ decays will then be very useful.
Moreover, even if sufficiently many different $\tilde \chi_i^0
\tilde \chi_j^0$ states are accessible to determine all the
parameters appearing in Eq.~(1), analyses of neutralino decay will
offer valuable redundancy. After all, a theory can only be said to
be tested successfully if experiments over--constrain its
parameters.

In the present work we systematically investigate, both analytically and
numerically, the usefulness of electron and positron beam polarization,
including transverse polarization, for the analysis of neutralino production
and decay at $e^+e^-$ colliders. To this end, we exploit spin/angular
correlations of the neutralino production $e^+ e^- \to \tilde{\chi}^0_2
\tilde{\chi}^0_1$ and subsequent two--body decays of $\tilde{\chi}^0_2 \to
\tilde{\chi}^0_1 h, \tilde{\chi}^0_1 Z,$ and $\tilde{\chi}^0_2 \to
\tilde{\ell}^\pm \ell^\mp$ followed by $\tilde{\ell}^\pm \to \ell^\pm
\tilde{\chi}^0_1$ for probing the CP properties of the neutralino sector in
the MSSM.  Due to the Majorana nature of neutralinos, the decay distributions
of two--body decays $\tilde{\chi}^0_2\to\tilde{\chi}^0_1  h, \tilde{\chi}^0_1 Z$ are
independent of the $\tilde{\chi}^0_2$ polarization, unless the polarization
of the $Z$ boson is measured. These modes can still be used to probe a
production--level CP--odd asymmetry, which however turns out to be small in
the MSSM. The slepton mode $\tilde{\chi}^0_2\to\tilde{\ell}^\pm_R\ell^\mp$ is
an optimal polarization analyzer of the decaying neutralino.
We can construct several CP--odd ``decay'' asymmetries that are sensitive to the
$\tilde \chi_2^0$ polarization vector. Our main emphasis is on observables
that {\em fully} reflect the non--trivial angular dependence of CP--odd terms,
except for the angular dependence appearing in the propagators. Although they
are not perfectly optimal, these CP--odd asymmetries have much higher
statistical significance than the conventional ones, as demonstrated with
numerical examples below.

The remainder of this article is organized as follows.  Section~\ref{sec:sec2}
describes neutralino production, including the polarization of the
neutralinos, for arbitrary beam polarization. Two--body decays of polarized
neutralinos are discussed in Sec.~\ref{sec:sec3}. Section~\ref{sec:sec4} deals
with the reconstruction of $\tilde \chi_1^0 \tilde \chi_2^0$ final states with
invisible $\tilde \chi_1^0$. The formalism of ``effective asymmetries'' is
described in Sec.~\ref{sec:sec5}, and numerical examples for these asymmetries
are shown in Sec.~\ref{sec:sec6}.  Finally, Section~\ref{sec:sec7} contains a
brief summary and some conclusions.

%%%%%%%%%%%%%%%%%%%%%%%%%%%%%%%%%%%%%%%%%%%%%%%%%%%%%%%%%%%%%%%%%%
\section{Neutralino production in \boldmath{$e^+e^-$} collisions}
\label{sec:sec2}
\setcounter{footnote}{0}
%%%%%%%%%%%%%%%%%%%%%%%%%%%%%%%%%%%%%%%%%%%%%%%%%%%%%%%%%%%%%%%%%%

\begin{figure}
{\color{blue}
\begin{center}
\begin{picture}(330,100)(10,0)
%1st diagram
\Text(5,85)[]{$e^-$}
\ArrowLine(0,75)(25,50)
\ArrowLine(25,50)(0,25)
\Text(5,15)[]{$e^+$}
\Photon(25,50)(65,50){2}{6}
\Text(45,37)[]{\color{red} $Z$}
\Line(65,50)(90,75)
\Photon(65,50)(90,75){2}{6}
\Text(87,87)[]{$\tilde{\chi}^0_i$}
\Line(90,25)(65,50)
\Photon(90,25)(65,50){2}{6}
\Text(87,13)[]{$\tilde{\chi}^0_j$}
% 2rd Diagram
\Text(125,85)[]{$e^-$}
\ArrowLine(120,75)(165,75)
\Text(125,15)[]{$e^+$}
\ArrowLine(165,25)(120,25)
\Line(164,75)(164,25)
\Line(166,75)(166,25)
\Text(150,50)[]{\color{red} $\tilde{e}_{L,R}$}
\Line(165,75)(210,75)
\Photon(165,75)(210,75){2}{6}
\Text(207,87)[]{$\tilde{\chi}^0_i$}
\Line(210,25)(165,25)
\Photon(210,25)(165,25){2}{6}
\Text(207,13)[]{$\tilde{\chi}^0_j$}
% 3rd Diagram
\Text(245,85)[]{$e^-$}
\ArrowLine(240,75)(285,75)
\Text(245,15)[]{$e^+$}
\ArrowLine(285,25)(240,25)
\Line(284,75)(284,25)
\Line(286,75)(286,25)
\Text(270,50)[]{\color{red} $\tilde{e}_{L,R}$}
\Line(285,75)(330,25)
\Photon(285,75)(330,25){2}{8}
\Text(327,87)[]{$\tilde{\chi}^0_i$}
\Line(330,75)(285,25)
\Photon(330,75)(285,25){2}{8}
\Text(327,13)[]{$\tilde{\chi}^0_j$}
\end{picture}\\
\end{center}
}
%%%%%\noindent
\caption{\it Feynman diagrams for five mechanisms contributing to the
  production of diagonal and non--diagonal neutralino pairs in $e^+e^-$
  annihilation, $e^+e^-\rightarrow \tilde{\chi}^0_i \tilde{\chi}^0_j$
  $(i,j$=1--4).}
\label{fig:diagrams}
\end{figure}
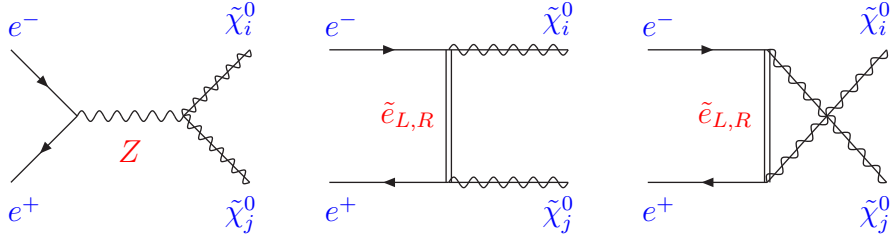

The neutralino pair production processes in $e^+e^-$ collisions
\begin{eqnarray} \label{process}
  e^-(p,\sigma) + e^+(\bar{p},\bar{\sigma})
\rightarrow
   \tilde{\chi}^0_i(p_i,\lambda_i)+\tilde{\chi}^0_j(p_j,\lambda_j)
   \qquad (\, i,j=1,2,3,4)
\end{eqnarray}
are generated by the five mechanisms of the Feynman diagrams in
Fig.\,\ref{fig:diagrams}, with $s$--channel $Z$ exchange, or $t$-- or
$u$--channel $\tilde{e}_{L,R}$ exchange. Here $\sigma$, $\bar{\sigma}$,
$\lambda_i$, and $\lambda_j$ denote helicities. For the analytical
calculation, we take a coordinate system where the production occurs in the
$(x,z)$ plane and the incident electron beam moves into $+z$ direction. The
four--momenta appearing in Eq.~(\ref{process}) are then given by
\begin{eqnarray} \label{mom}
p       &=& \frac{\sqrt{s}}{2} (1,\,0,\,0,\phantom{-}1) \, , \nonumber\\
\bar{p} &=& \frac{\sqrt{s}}{2} (1,\,0,\,0,-1) \, ,\nonumber\\
p_i     &=& \frac{\sqrt{s}}{2} (e_i, \phantom{-}\lambda^{1/2}\sin\Theta,\, 0,
                          \phantom{-}\lambda^{1/2}\cos\Theta) \,,\nonumber\\
p_j     &=& \frac{\sqrt{s}}{2} (e_j, -\lambda^{1/2}\sin\Theta,\, 0,
                                -\lambda^{1/2}\cos\Theta)\, ,
\end{eqnarray}
where
\begin{eqnarray} \label{mu}
&& e_i = 1+\mu_i^2 -\mu_j^2\, , \quad
\quad e_j = 1+\mu_j^2 -\mu_i^2 \, , \nonumber\\
&& \mu_{i,j} = m_{\tilde{\chi}^0_{i,j}}/\sqrt{s}\, ,\,\, \quad   \quad
   \lambda   = (1-\mu_i^2-\mu_j^2)^2-4\mu_i^2\mu_j^2 \, .
\end{eqnarray}

The transition matrix element, after an appropriate Fierz transformation of
the $\tilde{e}_{L,R}$ exchange amplitudes, can be expressed in terms of four
generalized bilinear charges $Q_{\alpha\beta}$:
\begin{eqnarray}
T\left(e^+e^-\rightarrow\tilde{\chi}^0_i\tilde{\chi}^0_j\right)
 = \frac{e^2}{s}\, Q_{\alpha\beta}
   \left[\bar{v}(e^+)  \gamma_\mu P_\alpha  u(e^-)\right]
   \left[\bar{u}(\tilde{\chi}^0_i) \gamma^\mu P_\beta
               v(\tilde{\chi}^0_j)\right] \, .
\label{amplitude}
\end{eqnarray}
These generalized charges correspond to independent helicity amplitudes which
describe the neutralino production processes for completely (longitudinally)
polarized electrons and positrons, neglecting the electron mass as well as
$\tilde e_L$--$\tilde e_R$ mixing.\footnote{$\tilde f_L$--$\tilde f_R$ mixing
  is proportional to $m_f$ unless one tolerates deeper minima of the scalar
  potential where charged sfermion fields obtain nonvanishing vacuum expectation
  values; although it can be enhanced at large $\tan\beta$
  or for large trilinear $A-$parameters, selectron mixing is generally
  negligible for collider physics purposes.} They are defined in terms of the
lepton and neutralino couplings as well as the propagators of the exchanged
(s)particles~\cite{CSS,Choi:2001ww}:
\begin{eqnarray} \label{bilinear}
&& Q_{LL}=+\frac{D_Z}{s_W^2c_W^2}\,
           (s_W^2-{\textstyle{\frac{1}{2}}}\,) {\cal Z}_{ij}
            -D_{uL}g_{Lij},\nonumber\\
&& Q_{RL}=+\frac{D_Z}{c_W^2}\,
            {\cal Z}_{ij}
            +D_{tR}g_{Rij},\nonumber\\
&& Q_{LR}=-\frac{D_Z}{s_W^2c_W^2}\,
           (s_W^2 -{\textstyle{\frac{1}{2}}}\,){\cal Z}^*_{ij}
            +D_{tL}g^*_{Lij},\nonumber\\
&& Q_{RR}=-\frac{D_Z}{c_W^2}{\cal Z}^*_{ij}
            -D_{uR}g^*_{Rij}.
\end{eqnarray}
The first index in $Q_{\alpha\beta}$ refers to the chirality of the $e^\pm$
current, the second index to the chirality of the $\tilde{\chi}^0$ current.
The first term in each bilinear charge is generated by $Z$--exchange and the
second term by selectron exchange; $D_Z$, $D_{tL,R}$ and $D_{uL,R}$
respectively denote the $s$--channel $Z$ propagator and the $t$-- and
$u$--channel left/right--type selectron propagators:
\begin{eqnarray}
&& D_Z=\frac{s}{s-m^2_Z+im_Z\Gamma_Z}\, ,\nonumber\\
&& D_{tL,R}=\frac{s}{t-m^2_{\tilde{e}_{L,R}}}  \qquad {\rm and }
   \qquad t\rightarrow u\, ,
\end{eqnarray}
with $s=(p + \bar p)^2$, $t=(p - p_i)^2$ and $u=(p - p_j)^2$. The matrices
${\cal Z}_{ij}$, $g_{Lij}$ and $g_{Rij}$ can be computed from the matrix $N$
diagonalizing the neutralino mass matrix \cite{book}
\begin{eqnarray}
&& {\cal Z}_{ij}= (N_{i3}N^*_{j3}-N_{i4}N^*_{j4})/2\, ,\nonumber\\
&& g_{Lij}=(N_{i2}c_W+N_{i1}s_W)(N^*_{j2}c_W+N^*_{j1}s_W)/
                4 s_W^2c_W^2 \, ,\nonumber\\
&& g_{Rij}=N_{i1}N^*_{j1}/c_W^2\, .
\label{eq:combinations}
\end{eqnarray}
They satisfy the hermiticity relations of
\begin{eqnarray}
{\cal Z}_{ij}={\cal Z}^*_{ji}\, , \qquad
g_{Lij}=g^*_{Lji}\, , \qquad g_{Rij}=g^*_{Rji}\, .
\end{eqnarray}
If the decay width $\Gamma_Z$ is neglected in the $Z$ boson propagator $D_Z$,
the bilinear charges $Q_{\alpha\beta}$ satisfy similar relations,
$Q_{\alpha\beta} (\neu_i, \neu_j, t,u) =Q_{\alpha\beta}^* (\neu_j, \neu_i,
u,t)$. These relations are very useful in classifying CP--even and CP--odd
observables.

%---------------------------------------------------
\subsection{Production helicity amplitudes}
%---------------------------------------------------

With the $e^\pm$ mass neglected, the matrix element in Eq.~(\ref{amplitude}) is
nonzero only if the electron helicity is opposite to the positron helicity.
We write the helicity amplitudes as
\begin{eqnarray} \label{chiral}
T(\sigma,\bar{\sigma},\lambda_i,\lambda_j) =
T(\sigma,-\sigma,\lambda_i,\lambda_j)\,\, \delta_{\bar{\sigma}, -\sigma}
 \equiv 2\pi \alpha\, \langle \sigma;\lambda_i\,\lambda_j \rangle\,\,
 \delta_{\bar{\sigma},-\sigma} \, ,
\end{eqnarray}
where $\sigma, \lambda_i, \lambda_j=\pm$.  Explicit expressions for these
helicity amplitudes are~\cite{CSS}:
\begin{eqnarray}
\langle +; ++ \rangle \!\!\!&=&\!\!\!
 -\left[Q_{RR} \sqrt{\eta_{i+}\eta_{j-}}
       +Q_{RL} \sqrt{\eta_{i-}\eta_{j+}}\right]
       \sin\Theta    \, , \nonumber \\
\langle +; +- \rangle \!\!\!&=&\!\!\!
 -\left[Q_{RR} \sqrt{\eta_{i+}\eta_{j+}}
       +Q_{RL} \sqrt{\eta_{i-}\eta_{j-}}\right]
       (1+\cos\Theta) \, , \nonumber \\
\langle +; -+ \rangle \!\!\!&=&\!\!\!
 +\left[Q_{RR} \sqrt{\eta_{i-}\eta_{j-}}
       +Q_{RL} \sqrt{\eta_{i+}\eta_{j+}}\right]
       (1-\cos\Theta)\, , \nonumber\\
\langle +; -- \rangle \!\!\!&=&\!\!\!
 +\left[Q_{RR} \sqrt{\eta_{i-}\eta_{j+}}
       +Q_{RL} \sqrt{\eta_{i+}\eta_{j-}}\right]
       \sin \Theta \, ,   \nonumber\\
\langle -; ++ \rangle \!\!\!&=&\!\!\!
 -\left[Q_{LL} \sqrt{\eta_{i-}\eta_{j+}}
       +Q_{LR} \sqrt{\eta_{i+}\eta_{j-}}\right]
       \sin \Theta \, ,   \nonumber\\
\langle -; +- \rangle \!\!\!&=&\!\!\!
 +\left[Q_{LL} \sqrt{\eta_{i-}\eta_{j-}}
       +Q_{LR} \sqrt{\eta_{i+}\eta_{j+}}\right]
       (1-\cos\Theta)\, , \nonumber\\
\langle -; -+ \rangle \!\!\!&=&\!\!\!
 -\left[Q_{LL} \sqrt{\eta_{i+}\eta_{j+}}
       +Q_{LR} \sqrt{\eta_{i-}\eta_{j-}}\right]
       (1+\cos\Theta) \, , \nonumber\\
\langle -; -- \rangle \!\!\!&=&\!\!\!
 +\left[Q_{LL} \sqrt{\eta_{i+}\eta_{j-}}
       +Q_{LR} \sqrt{\eta_{i-}\eta_{j+}}\right]
       \sin \Theta\, ,
\label{renamp}
\end{eqnarray}
where $\eta_{i\pm} = e_i\pm \lambda^{1/2}$ and $\eta_{j\pm}=e_j\pm
\lambda^{1/2}$. In the high energy asymptotic limit, $\eta_{i+}$ and
$\eta_{i-}$ approach 1 and 0, respectively; only the
helicity amplitudes with opposite $\tilde{\chi}^0_i$ and
$\tilde{\chi}^0_j$ helicities survive.

%---------------------------------------------------
\subsection{Production cross sections}
%---------------------------------------------------

We analyze neutralino production for general $e^\pm$ polarization
states. With the scattering plane fixed as the $(x,z)$
plane, the azimuthal scattering angle appears in the description of
the $e^\pm$ polarization vectors:
\begin{eqnarray} \label{polvec}
\overrightarrow{P}_{e^-} = (P_T \cos\Phi, -P_T \sin\Phi, P_L),
    \quad
\overrightarrow{P}_{e^+} = (\overline{P}_T \cos(\eta-\Phi),
                    \overline{P}_T \sin(\eta-\Phi), -\overline{P}_L) \, ,
\end{eqnarray}
where $\eta$ is the relative angle between the transverse components of two
polarization vectors. The density matrices $\rho$ ($\overline{\rho}$) of the
electron (positron)
in the $\{+,-\}$ helicity basis are \cite{Hagiwara:1985yu}
\begin{eqnarray} \label{rho}
\rho
  = \frac{1}{2} \left(\begin{array}{cc}
                 1+P_L & P_T e^{i \Phi} \\
                 P_T e^{-i \Phi} & 1-P_L
                      \end{array}\right), \qquad
\overline{\rho}
  = \frac{1}{2} \left(\begin{array}{cc}
      1+\overline{P}_L & -\overline{P}_T e^{-i(\eta-\Phi)}\\
      -\overline{P}_T e^{i(\eta-\Phi)}  & 1-\overline{P}_L \\
                      \end{array}\right) \, .
\end{eqnarray}

The polarized differential cross section is given by
\begin{eqnarray}
\frac{d \sigma}{d\Omega}
   = \frac{\lambda^{1/2}}{64\pi^2 s} \overline{|T|^2} \, ,
\end{eqnarray}
where
\begin{eqnarray} \label{tsq}
\overline{|T|^2}
  = \sum_{\sigma,\bar{\sigma},\lambda_i,\lambda_j}\,
         T(\sigma,\bar{\sigma},\lambda_i,\lambda_j)
         T^*(\sigma',\bar{\sigma}',\lambda_i,\lambda_j)\,
         \rho_{\sigma\sigma'}\,
         \overline{\rho}_{\bar{\sigma}'\,\bar{\sigma}} \, .
\end{eqnarray}
Note that the order of indices of $\overline{\rho}_{\bar{\sigma}' \,
  \bar{\sigma}}$ is opposite of that of $\rho_{\sigma\sigma'}$ due to the
difference between the particle and the antiparticle. Inserting
Eqs.~(\ref{renamp}) and (\ref{rho}) into Eq.~(\ref{tsq}) yields
\begin{eqnarray}
&& \frac{d \sigma}{d \Omega}\{ij\}
  =\frac{\alpha^2}{4 s}\, \lambda^{1/2} \bigg[
     (1-P_L\bar{P}_L)\,\Sigma^{ij}_{UU}+(P_L-\bar{P}_L)\,\Sigma^{ij}_{UL}
     \nonumber\\
&& { }\hskip 3cm
  +P_T\bar{P}_T\cos(2\Phi-\eta)\,\Sigma^{ij}_{UT}
  +P_T\bar{P}_T\sin(2\Phi-\eta)\,\Sigma^{ij}_{UN}\bigg] \, ,
\label{diffx}
\end{eqnarray}
where
\begin{eqnarray}
&& \Sigma^{ij}_{UU}=\left[1-(\mu^2_i - \mu^2_j)^2
                   +\lambda\cos^2\Theta\right]Q_1
                   +4\mu_i\mu_j Q_2+2\lambda^{1/2} Q_3\cos\Theta,
                  \nonumber\\
&& \Sigma^{ij}_{UL}=\left[1-(\mu^2_i - \mu^2_j)^2
                   +\lambda\cos^2\Theta\right]Q'_1
                   +4\mu_i\mu_j Q'_2+2\lambda^{1/2} Q'_3\cos\Theta,
                  \nonumber\\
&& \Sigma^{ij}_{UT}= \lambda \, Q_5 \sin^2\Theta, \nonumber\\
&& \Sigma^{ij}_{UN}=-\lambda \, Q'_6 \sin^2\Theta \, .
\label{SUB}
\end{eqnarray}
Expressions for all relevant quartic charges $Q_i^{(\prime)}$
in terms of bilinear charges $Q_{\alpha\beta}$ are
given in Table~\ref{tab:quartic}, which also lists the transformation
properties under P and CP. Non--zero transverse $e^\pm$ beam polarization
allows to probe four new quartic charges, $Q_5$, $Q_6$, $Q'_5$, and $Q'_6$.

\begin{table*}[\hbt]
\caption[{\bf Table 1:}]{\label{tab:quartic}
{\it The independent quartic charges describing $e^+e^- \to \tilde \chi_i^0
  \tilde \chi_j^0$.}}
\begin{center}
\begin{tabular}{|c|c|l|}\hline
 &  &  \\[-4mm]
${\rm P}$ & ${\rm CP}$ & { }\hskip 2cm Quartic charges \\\hline \hline
 &  &  \\[-3mm]
 even    &  even     & $Q_1 =\frac{1}{4}\left[|Q_{RR}|^2+|Q_{LL}|^2
                       +|Q_{RL}|^2+|Q_{LR}|^2\right]$ \\[2mm]
         &           & $Q_2 = \frac{1}{2}\real\left[Q_{RR}Q^*_{RL}
                       +Q_{LL}Q^*_{LR}\right]$ \\[2mm]
         &           & $Q_3 = \frac{1}{4}\left[|Q_{RR}|^2+|Q_{LL}|^2
                       -|Q_{RL}|^2-|Q_{LR}|^2\right]$ \\[2mm]
         &           & $Q_5=\frac{1}{2}\real \left[Q_{RR}Q^*_{LR}
                       +Q_{LL}Q^*_{RL}\right]$ \\
 & & \\[-3mm]
\cline{2-3}
 & & \\[-3mm]
         &  odd      & $Q_4=\frac{1}{2}\imag\left[Q_{RR}Q^*_{RL}
                       +Q_{LL}Q^*_{LR}\right]$\\[2mm]
         &           & $Q_6=\frac{1}{2}\imag\left[Q_{RR}Q^*_{LR}
                       +Q_{LL}Q^*_{RL}\right]$ \\[2mm]\hline \hline
 & & \\[-3mm]
 odd     &  even     & $Q'_1=\frac{1}{4}\left[|Q_{RR}|^2+|Q_{RL}|^2
                        -|Q_{LL}|^2-|Q_{LR}|^2\right]$\\[2mm]
         &           & $Q'_2=\frac{1}{2}\real\left[Q_{RR}Q^*_{RL}
                        -Q_{LL}Q^*_{LR}\right]$ \\[2mm]
         &           & $Q'_3=\frac{1}{4}\left[|Q_{RR}|^2+|Q_{LR}|^2
                        -|Q_{LL}|^2-|Q_{RL}|^2\right]$\\[2mm]
         &           & $Q'_5=\frac{1}{2}\real\left[Q_{RR}Q^*_{LR}
                        -Q_{LL}Q^*_{RL}\right]$ \\
 & & \\[-3mm]
\cline{2-3}
 & & \\[-3mm]
         & odd       & $Q'_4=\frac{1}{2}\imag\left[Q_{RR}Q^*_{RL}
                       -Q_{LL}Q^*_{LR}\right]$
                       \\[2mm]
         &           & $Q'_6=\frac{1}{2}\imag\left[Q_{RR}Q^*_{LR}
                       -Q_{LL}Q^*_{RL}\right]$ \\[2mm]
\hline
\end{tabular}
\end{center}
\end{table*}

\subsection{Neutralino polarization vector}
\label{sec:pol}

%Even if the initial beams are not polarized, the chiral structure
%of the neutralinos could be inferred from the polarization of the
%$\tilde{\chi}^0_i\tilde{\chi}^0_j$ pairs produced in $e^+e^-$
%annihilation. However, the neutralino polarization does depend on the
%polarization of the initial state.

The polarization vector $\vec{\cal P}^i=({\cal P}_T^i, {\cal P}_N^i,{\cal
  P}_L^i) $ of the neutralino $\tilde{\chi}^0_i$ is defined in its rest frame.
The longitudinal component ${\cal P}_L^i$ is parallel to the
$\tilde{\chi}^0_i$ flight direction in the c.m. frame, ${\cal P}_T^i$ is in
the production plane, and ${\cal P}_N^i$ is normal to the production plane.
In order to extract the vector $\vec{\cal P}^i$, we first define the
polarization density matrix for the out--going neutralino $\tilde{\chi}^0_i$:
\begin{eqnarray}
\rho^i_{\lambda_i \lambda_i'} = \frac{\sum_{\sigma,\lambda_j}\,
                                \langle \sigma;\lambda_i\lambda_j\rangle
                                \langle \sigma;\lambda'_i\lambda_j\rangle^*}{
           \sum_{\sigma,\lambda_i,\lambda_j}\,
                              \langle \sigma;\lambda_i\lambda_j\rangle
                              \langle \sigma;\lambda_i\lambda_j\rangle^* }\, .
\end{eqnarray}
Explicit expressions for the helicity amplitudes $\langle \sigma; \lambda_i
\lambda_j \rangle$ are given in Eq.~(\ref{renamp}). The polarization vector
of the neutralino $\tilde{\chi}^0_i$ is then given by
\begin{eqnarray} \label{pdel}
\vec{\cal P}^i = \mbox{Tr} ( \overrightarrow{\sigma} \rho^i)
  = \frac{1}{\Delta^{ij}_U}\left(\Delta^{ij}_T, \Delta^{ij}_N,
             \Delta^{ij}_L\right) \, .
\end{eqnarray}
We can decompose the three polarization components as well as the unpolarized
part according to combinations of $e^\pm$ polarizations:
\begin{eqnarray}
 \Delta^{ij}_U \!\!\!&=&\!\!\!
          (1-P_L \overline{P}_L) \Sigma^{ij}_{UU}
          +(P_L-\overline{P}_L) \Sigma^{ij}_{UL}
          + P_T \overline{P}_T \{\Sigma^{ij}_{UT}c_{(2\Phi-\eta)}
                                 +\Sigma^{ij}_{UN} s_{(2\Phi-\eta)}\} \, ,
          \nonumber \\[1mm]
 \Delta^{ij}_L \!\!\!&=&\!\!\!
          (1-P_L \overline{P}_L) \Sigma^{ij}_{LU}
          \,+(P_L-\overline{P}_L) \Sigma^{ij}_{LL}
          + P_T \overline{P}_T \{\Sigma^{ij}_{LT}c_{(2\Phi-\eta)}
                                 +\Sigma^{ij}_{LN} s_{(2\Phi-\eta)}\} \, ,
          \nonumber \\[1mm]
 \Delta^{ij}_T \!\!\!&=&\!\!\!
          (1-P_L \overline{P}_L) \Sigma^{ij}_{TU}
          \,+(P_L -\overline{P}_L) \Sigma^{ij}_{TL}
          + P_T \overline{P}_T \{\Sigma^{ij}_{TT} c_{(2\Phi-\eta)}
                                +\Sigma^{ij}_{TN} s_{(2\Phi-\eta)}\} \, ,
          \nonumber\\[1mm]
 \Delta^{ij}_N \!\!\!&=&\!\!\!
          (1-P_L \overline{P}_L) \Sigma^{ij}_{NU}
          + (P_L -\overline{P}_L) \Sigma^{ij}_{NL}
          + P_T \overline{P}_T \{\Sigma^{ij}_{NT} c_{(2\Phi-\eta)}
                                \!+\Sigma^{ij}_{NN} s_{(2\Phi-\eta)}\} \, ,
\label{polcomp}
\end{eqnarray}
where $c_{(2\Phi-\eta)}=\cos (2\Phi-\eta)$, $s_{(2\Phi-\eta)} =
\sin(2\Phi-\eta)$, and the $\Sigma_{UB} \ ( B=U,\, L,\, T, \, N)$ are in
Eq.~(\ref{SUB}). The $\Sigma_{BU}$, which survive even without beam
polarization, are given by
{\small
\begin{eqnarray} \label{SBU}
\Sigma^{ij}_{LU} \!\!\!&=&\!\!\!\phantom{-}
               2(1-\mu^2_i-\mu^2_j)\,\cos\Theta\,Q'_1
              +4\mu_i\mu_j\,\cos\Theta\, Q'_2
              +\lambda^{1/2}\{1+\cos^2\Theta
              -\sin^2\Theta(\mu^2_i-\mu^2_j)\}\, Q'_3 \, ,
              \nonumber \\
\Sigma^{ij}_{TU}\!\!\!&=&\!\!\!-2\sin\Theta\left[\{(1-\mu^2_i+\mu^2_j)\,Q'_1
              +\lambda^{1/2} \cos\Theta \, Q'_3\}\mu_i
              +(1+\mu^2_i-\mu^2_j)\mu_j\,Q'_2\right] \, , \nonumber\\
\Sigma^{ij}_{NU}\!\!\!&=&\!\!\!\phantom{-}2\lambda^{1/2}\mu_j\,\sin\Theta\, Q_4
\, .
\end{eqnarray}
}
\hskip -0.2cm The remaining $\Sigma_{AB}$,
which contribute only with non--trivial $e^\pm$ polarization, are
\begin{eqnarray} \label{SAB}
\Sigma^{ij}_{LL} \!\!\!
     &=&\!\!\! \phantom{-}
        [\lambda+1-(\mu^2_i-\mu^2_j)^2]\,\cos\Theta\,Q_1
        +4\mu_i\mu_j\,\cos\Theta\, Q_2 \nonumber\\
       &&\hskip 1.3cm
     +\lambda^{1/2} [1 +\cos^2\Theta - {\sin^2\!\Theta} \, (\mu^2_i -
     \mu^2_j)] \, Q_3 \, ,
        \nonumber \\
 \Sigma^{ij}_{LT} \!\!\! &=& \!\!\! \phantom{-} \lambda^{1/2} ( 1 + \mu^2_i -
     \mu^2_j) \, {\sin^2\Theta} \, Q'_5 \, ,
        \nonumber \\
 \Sigma^{ij}_{LN} \!\!\!
     &=&\!\!\! - \lambda^{1/2}(1+\mu^2_i-\mu^2_j)\,{\sin^2\Theta}\,Q_6 \, ,
        \nonumber \\
 \Sigma^{ij}_{TL}\!\!\!
     &=&\!\!\! -2\sin\Theta\left\{[(1-\mu^2_i+\mu^2_j)\,Q_1
              +\lambda^{1/2}\cos\Theta\, Q_3]\mu_i
              +(1+\mu^2_i-\mu^2_j)\mu_j\,Q_2\right\} \, ,
        \nonumber\\
\Sigma^{ij}_{TT} \!\!\!&=&\!\!\! \phantom{-} \lambda^{1/2} \mu_i \sin 2 \Theta
        \, Q'_5 \, ,
        \nonumber\\
 \Sigma^{ij}_{TN} \!\!\!&=&\!\!\! -\lambda^{1/2}\mu_i\sin2 \Theta \,Q_6 \, ,
        \nonumber\\
\Sigma^{ij}_{NL} \!\!\!&=&\!\!\! \phantom{-} 2 \lambda^{1/2} \mu_j \,
        \sin\Theta\, Q'_4 \, ,
        \nonumber\\
\Sigma^{ij}_{NT} \!\!\!&=&\!\!\! -2\lambda^{1/2}\mu_i\,\sin\Theta\, Q_6 \, ,
        \nonumber\\
\Sigma^{ij}_{NN} \!\!\!&=&\!\!\! -2\lambda^{1/2}\mu_i\,\sin\Theta\, Q'_5 \, .
\end{eqnarray}
The P and CP properties of all these quantities are identical to
those of the quartic charges in Table~\ref{tab:quartic}. In
particular, the five quantities $\Sigma_{UN}, \Sigma_{LN},
\Sigma_{TN}, \Sigma_{NU}$ and $\Sigma_{NL}$ are CP--odd.

Brief comments on the reference frame are in order here.  In the coordinate
system which we have employed so far, the scattering plane is fixed, while the
direction of $e^\pm$ transverse polarization vectors differs from event to
event. For a real experiment, fixed $e^\pm$ polarization vectors should be
more convenient. We define the transverse part of $\vec P_{e^-}$ as $+x$
direction; the $x$ and $y$ components of the outgoing neutralino
four--momentum $p_i$ are then proportional to $\cos\Phi$ and $\sin\Phi$,
respectively. In this coordinate system the scattering plane changes from
event to event. Since only the {\rm relative} angles between the $e^\pm$
polarization vectors and the scattering plane are relevant, the final results
in Eqs.~(\ref{diffx}) and (\ref{polcomp}) are still valid. In this new coordinate
frame, the $\tilde \chi_i^0$ polarization vector can be explicitly written as
\begin{equation} \label{polvec1}
\vec{\cal P}^i = {\cal P}_T^i \vec e_T + {\cal P}_N^i \vec e_N +
{\cal P}_L^i \vec e_L\, ,
\end{equation}
where the following three unit vectors
form a co--moving orthonormal basis of the three--dimensional space:
\begin{eqnarray} \label{polvec2}
\vec e_T &=& (\cos \Phi \cos \Theta, \, \sin \Phi \cos \Theta, \, -\sin
\Theta) \, , \nonumber \\
\vec e_N &=& (-\sin \Phi, \, \cos \Phi, \, 0) \, , \nonumber \\
\vec e_L &=& (\cos \Phi \sin \Theta, \, \sin \Phi \sin \Theta, \, \cos
\Theta) \, .
\end{eqnarray}

Probing CP violation in the MSSM neutralino sector involves the four quartic
charges $Q_4, Q'_4, Q_6$ and $Q'_6$ for $i\neq j$.  Their characteristic
features can be analytically understood from their explicit expressions in
terms of the neutralino mixing matrix $N$. With $\Gamma_Z$ neglected in
the high energy limit, they are
\begin{eqnarray}
Q^{(\prime)}_4
  &=& \frac{1}{2c^4_W s^4_W}\left[s^4_W\mp(s^2_W-1/2)^2\right] D^2_Z
    \imag({\cal Z}^2_{ij}) \nonumber\\
  && +\frac{D_Z}{2c^2_W}\left[(D_{tR}+D_{uR})\imag({\cal Z}_{ij} g_{Rij})
   \pm \frac{s^2_W-1/2}{s^2_W}(D_{tL}+D_{UL})\imag({\cal Z}_{ij} g_{Lij})\right]
     \nonumber\\
  && +\frac{1}{2} D_{uR}D_{tR}\imag(g^2_{Rij})
 \mp\frac{1}{2} D_{uL}D_{tL}\imag(g^2_{Lij})\, ,
   \nonumber\\[1mm]
Q^{(\prime)}_6 &=& \frac{1}{2c^2_W}D_Z(D_{tL}\pm D_{uL})
     \imag({\cal Z}_{ij} g^*_{Lij})
     +\frac{s^2_W-1/2}{2s^2_W c^2_W} D_Z (D_{uR}\pm D_{tR})
     \imag({\cal Z}_{ij} g^*_{Rij})\nonumber\\
  && + \frac{1}{2}(D_{uR}D_{tL}\pm D_{tR}D_{uL})
     \imag(g_{Lij}g^*_{Rij}) \, ,
\label{quart}
\end{eqnarray}
where the explicit form of ${\cal Z}_{ij}$, $g_{Lij}$ and $g_{Rij}$ are listed
in Eq.~(\ref{eq:combinations}). From the propagator combinations, we see that
the quartic charge $Q'_6$ is forward--backward asymmetric with respect to the
scattering angle $\Theta$ while the other three quartic charges,
$Q^{(\prime)}_4$ and $Q_6$, are forward--backward symmetric.

The relative sizes of the four CP--violating quartic charges indicate which
observables should be promising to investigate experimentally.  Let us first
consider the generic case of small gaugino--higgsino mixing (with substantial
CP phase $\Phi_1$). Small mixing is generally obtained if the entries in the
off--diagonal $2 \times 2$ blocks in the neutralino mass matrix are
smaller than those in the diagonal blocks, allowing an expansion in powers of
$m_Z$. Analytic expressions for $N$ using this expansion, given in
Ref.~\cite{Choi:2004rf}, help to estimate the sizes of the Q
$Q_{4,6}^{(\prime)}$. In particular, the last term contributing to
$Q_4^{(\prime)}$ in Eq.~(\ref{quart}), which is proportional to $\sin\Phi_1$,
is not suppressed by small mixing angles: $Q_4$ and $Q'_4$ survive even
without any gaugino--higgsino mixing. In contrast $Q_6$ and $Q'_6$ only start
at $O(m^2_Z)$. This is related to the observation that, in the notation of
Ref.~\cite{Choi:2001ww}, $Q_6^{(\prime)}$ probe Dirac--type phases, which
vanish in the absence of nontrivial mixing between neutralino current
eigenstates, whereas $Q_4^{(\prime)}$ probe Majorana--type phases, which
survive in this limit. In the generic case of small gaugino--higgsino mixing,
therefore, the size of $Q_4^{(\prime)}$ is much larger than that of
$Q_6^{(\prime)}$. In the case of strong gaugino--higgsino mixing, however,
$Q_6^{(\prime)}$, which can only be probed with transversely polarized beams,
could exceed $Q_4$ and/or $Q'_4$.

%%%%%%%%%%%%%%%%%%%%%%%%%%%%%%%%%%%%%%%%%%%%%
\section{Two--body neutralino decays}
\label{sec:sec3}
\setcounter{footnote}{0}
%%%%%%%%%%%%%%%%%%%%%%%%%%%%%%%%%%%%%%%%%%%%%

The decay patterns of heavy neutralinos $(\neu_{i>1})$ depend on
their masses and the masses and couplings of other sparticles and
Higgs bosons. In this article we focus on the two--body decays of
neutralinos. It is possible that the kinematics prohibits some two--body
tree--level decays. However, a sufficiently heavy neutralino can
decay via tree--level two--body channels containing a $Z$ or a Higgs
boson and a lighter neutralino \cite{Choi:2003fs}, and/or into a
sfermion--matter fermion pair.

Of particular interest in the present work are the following
two--body decay modes:
\begin{eqnarray}
\tilde{\chi}^0_i\to\tilde{\chi}^0_k\, Z, \qquad
\tilde{\chi}^0_i\to\tilde{\chi}^0_k\, h \quad \mbox{and}\quad
\tilde{\chi}^0_i\to\tilde{\ell}^\pm_R \ell^\mp \,,
\label{eq:2-body_decays}
\end{eqnarray}
with $\ell = e$ or $\mu$. If any of these processes is kinematically allowed,
it will dominate any tree--level three--body decay.
% and any two--body
%decay which is forbidden at the tree level.

The relevant couplings are
\begin{eqnarray}
&& \langle\, \ell^-_L|\,\tilde{\ell}^-_R\,|\tilde{\chi}^0_i\, \rangle
 =+\langle\, \ell^+_L|\,\tilde{\ell}^+_R\,|\tilde{\chi}^0_i\, \rangle^*
       = -\sqrt{2} g t_W\, N^*_{i1},\qquad
 \langle \ell^\pm_R|\tilde{\ell}^\pm_R|\tilde{\chi}^0_i \rangle
       = 0\, , \\[2mm]
&& \langle \tilde{\chi}^0_{kR} | Z | \tilde{\chi}^0_{iR}\rangle
 =-\langle \tilde{\chi}^0_{kL} | Z | \tilde{\chi}^0_{iL}\rangle^*
 = +\frac{g}{2c_W} \left[N_{i3} N^*_{k3}-N_{i4} N^*_{k4}\right]\, ,
 \nonumber\\ \nonumber
&& \langle \tilde{\chi}^0_{kL}| h | \tilde{\chi}^0_{iR}\rangle
 =+\langle \tilde{\chi}^0_{kR}| h | \tilde{\chi}^0_{iL}\rangle^*
 = \frac{g}{2}\left[(N_{k2}-t_W N_{k1})(s_\alpha N_{i3}+c_\alpha N_{i4})
                          +(i\leftrightarrow k)\right]\, ,
\end{eqnarray}
where $s_\alpha=\cos\alpha$, $c_\alpha=\sin\alpha$, and $\alpha$ being the
mixing angle between the two CP--even Higgs states in the MSSM \cite{book}.
Note that the $Z$ coupling is proportional to the higgsino components of
both participating neutralinos, whereas the Higgs coupling requires a higgsino
component of one neutralino and a gaugino component of the
other.\footnote{If $\delta m_{\tilde \chi} \equiv m_{\tilde \chi_2^0}
- m_{\tilde \chi_1^0} \gg m_Z$, the decay into longitudinally polarized $Z$
bosons gets enhanced by a factor $(\delta m_{\tilde \chi} / m_Z)^2$.
If $\delta m_{\tilde \chi} \sim {\cal O}(m_Z)$, three--body decays $
\tilde \chi_2^0 \rightarrow \tilde \chi_1^0 f\bar{f}$ may dominate
over $\tilde \chi_2^0 \rightarrow \tilde \chi_1^0 Z$ decays if
$|\mu| \gg m_{\tilde f}$; this does not happen in models where the
entire sparticle spectrum is described by a small number of
parameters.} Since the
lighter neutralino states $\tilde \chi_{1,2}^0$ are often gaugino--like, this
pattern of couplings implies that $\tilde \chi_i^0 \to \tilde \chi_1^0 h$
decays will often dominate over the (kinematically preferred) $\tilde \chi_i^0
\to \tilde \chi_1^0 Z$ decays. However, the $\tilde{\chi}^0_i \to
\tilde{\ell}^\pm_R \ell^\mp$ decays only depend on the gaugino components of
the decaying neutralino. If kinematically accessible, they can
have the largest branching ratios.

Note also that the Majorana nature of neutralinos relates the left-- and
right--handed couplings of the $Z$ and $h$ boson to a neutralino pair; they
are complex conjugate to each other, having an identical absolute magnitude.
These relations lead to a characteristic property of the corresponding
two--body decays, $\tilde{\chi}^0_i\to \tilde{\chi}^0_k Z$ and
$\tilde{\chi}^0_i\to\tilde{\chi}^0_k h$: {\em the decay distributions are
  independent of the polarization of the decaying neutralino
  $\tilde{\chi}^0_i$, unless the polarization of the $Z$ boson  or
  $\tilde{\chi}_k^0$ is measured}. In contrast, the slepton mode
in Eq.$\,$(\ref{eq:2-body_decays}) can be
exploited as optimal polarization analyzer of the decaying neutralino, if
the small lepton mass is
ignored; as noted earlier, this implies that $\tilde \ell_L$--$\tilde \ell_R$
mixing is ignored as well.\footnote{$\tilde \chi_i \to \tilde \tau_1^\pm
  \tau^\mp$ decays, where $\tilde \tau_L$--$\tilde \tau_R$ mixing can be
  important, have been analyzed in Refs.~\cite{cdgs}.}

Furthermore, the decay distributions are completely determined by the relevant
particle masses, as well as by the $\tilde \chi_i^0$ polarization vector
(in case of $\tilde \chi_i^0 \rightarrow \tilde
\ell_R^\pm \ell^\mp$ decay). More
explicitly, the angular distribution in the rest frame of the
decaying neutralino $\tilde{\chi}^0_i$ is
\begin{eqnarray} \label{decdist}
\frac{1}{\Gamma_X}\frac{d\Gamma_X}{d\Omega^*}
 = \frac{1}{4\pi}\left(1\pm \xi_X \vec{\cal P}^i \cdot \hat{ k}_1^*
 \right)\, ,
\end{eqnarray}
where $\xi_{Z, h}=0$ for the $Z$ and $h$ decay modes, and $\xi_{l^\pm}=\mp 1$
for $\tilde{\chi}^0_i\to \tilde{\ell}^\pm_R \ell^\mp$ with $\hat{ k}_1^*$
being the unit vector in $\ell^\mp$ direction.
% in the $\tilde \chi_i^0$ rest frame. <-In the same sentence, we said
% in the rest frame..
The former two decay modes can probe only ``production''
asymmetries, whereas the (s)leptonic decay mode can probe ``decay''
asymmetries also, which are sensitive to the $\tilde \chi_i^0$
polarization.

%%%%%%%%%%%%%%%%%%%%%%%%%%%%%%%%%%%%%%
\section{Event reconstruction}
\label{sec:sec4}
\setcounter{footnote}{0}
%%%%%%%%%%%%%%%%%%%%%%%%%%%%%%%%%%%%%%

We focus on $e^+ e^- \to \tilde \chi_2^0 \tilde \chi_1^0$ production, and
assume $\tilde \chi_1^0$ to be stable (or possibly to decay invisibly). The
only visible final state particles therefore result from $\tilde \chi_2^0$
decay, which simplifies the analysis. Moreover, this is the kinematically most
accessible neutralino pair production with visible final state; indeed, it is
often the first sparticle production channel accessible at $e^+ e^-$ colliders
\cite{abdel}.

An important difference between $\tilde \chi_2^0 \to \tilde \chi_1^0 (h,Z)$
and $\tilde \chi_2^0 \to \tilde \ell_R^\pm \ell^\mp \to \tilde \chi_1^0 \ell^+
\ell^-$ is the degree of event reconstruction. The latter decay chain allows
complete event reconstruction (with an, at least, two--fold ambiguity),
whereas the former does not. This can be seen by counting unknowns. The
$\tilde \chi_1^0 \tilde \chi_1^0 (h,Z)$ final states contain six unknown
components of $\tilde \chi_1^0$ momenta (we are assuming that the masses of
all produced particles have already been determined \cite{rec}, so that the
energies can be computed from three--momenta); this has to be compared with
four constraints from energy--momentum conservation, and a single mass
constraint, $(p_{\tilde \chi_1^0} + p_{(h,Z)})^2 = m^2_{\tilde \chi_2^0}$. One
quantity remains undetermined.

In contrast, $\tilde \chi_1^0 \tilde \chi_1^0 \ell^+ \ell^-$ final states
produced from an on--shell $\tilde \ell_R^\pm$ have two invariant mass
constraints. With an equal number of constraints and unknowns, the event can
be reconstructed \cite{newbartl}. An explicit reconstruction may proceed as
follows. Let $k_1$ and $k_2$ be the four--momenta of the two charged leptons
in the final state, and $p_1$ and $q$ the four--momenta of the two
neutralinos; here $k_2$ and $q$ originate from $\tilde \ell_R$ decay. Note
that the energy $p_1^0$ is fixed from two--body kinematics, see
Eq.~(\ref{mom}). Then $q^0$ is determined from energy conservation, once
the lepton energies are measured. The invariant mass constraint $(k_2 + q)^2
= m^2_{\tilde \ell_R}$ can fix the scalar product $\vec k_2 \cdot \vec q$.
The second mass constraint $(k_1 + k_2 + q)^2 = m^2_{\tilde \chi_2^0}$ is
used for $\vec k_1 \cdot \vec q$.  When writing the unknown three--momentum
$\vec q$ as $\vec q = a \vec k_1 + b \vec k_2 + c (\vec k_1 \times \vec
k_2)$, the two coefficients $a$ and $b$ can be computed from the two scalar
products $\vec k_2 \cdot \vec q$ and $\vec k_1 \cdot \vec q$ determined
above; note that the term proportional to $c$ drops out here. The last
coefficient $c$ can be computed from the known energy $q^0$ with two--fold
ambiguity.

Once $\vec q$ is known, $\vec p_1$ follows immediately from momentum
conservation. We can read off the production angles $\Theta$ and $\Phi$. This
also allows to compute the $\tilde \chi_2^0$ three--momentum $\vec p_2 = \vec
k_1 + \vec k_2 + \vec q = - \vec p_1$ (in the c.m. frame). With the known
$\tilde \chi_2^0$ energy, we boost into the $\tilde \chi_2^0$ rest frame, and
read off the $\tilde \chi_2^0$ decay angles $\Theta^*$ and $\Phi^*$; recall
that there is a non--trivial dependence on these decay angles via
Eq.~(\ref{decdist}).

So far we have assumed that we know which of the two charged leptons in the
final state originates from the $\tilde \chi_2^0$ decay, and which one from
$\tilde \ell_R$ decay. Since, owing to its Majorana nature, $\tilde \chi_2^0$
will decay into both $\tilde \ell_R^+ \ell^-$ and $\tilde \ell_R^- \ell^+$
final states with equal branching ratios, the charge of the leptons does not
help this discrimination of the origin of two charged leptons.
A unique assignment is nevertheless possible if the
two mass differences $\delta_{2R} \equiv m_{\tilde \chi_2^0} - m_{\tilde
  \ell_R}$ and $\delta_{R1} \equiv m_{\tilde \ell_R} - m_{\tilde \chi_1^0}$
are very different from each other: if $\delta_{2R} \gg \delta_{R1}$, the more
energetic (harder) lepton will originate from the first step of $\tilde
\chi_2^0$ decay, and the less energetic (softer) lepton comes from $\tilde
\ell_R$ decay; if $\delta_{2R} \ll \delta_{R1}$ the opposite assignment holds.
However, if $\delta_{2R} \simeq \delta_{R1}$, both assignments often lead to
physical solutions if the procedure for event reconstruction outlined above is
applied. In this unfavorable situation there is a four--fold ambiguity in the
event reconstruction.

Finally, we note that background events can be also reconstructed, in some
cases again with two--fold ambiguity. The main backgrounds to $\tilde \chi_2^0
\to \tilde \chi_1^0 (Z,h)$ decays are $e^+e^- \rightarrow ZZ, \, Zh$
production with one $Z$ decaying invisibly.  The $e^+e^- \to ZZ (\to \nu \bar
\nu \ell^+ \ell^-), \ W^+ W^- (\to \ell^+ \nu_\ell \ell^- \bar \nu_\ell), \
\tilde \ell^+ \tilde \ell^- (\to \ell^+ \ell^- \tilde \chi_1^0 \tilde
\chi_1^0)$ are the main backgrounds to $\tilde \chi_1^0 \tilde \chi_2^0 \to
\ell^+ \ell^- \tilde \chi_1^0 \tilde \chi_1^0$ production.\footnote{Note that
  we include supersymmetric slepton production as background, since it does
  not contribute to the CP--odd asymmetries we wish to analyze here.} We can
obtain a pure sample of signal events by discarding all events that can be
reconstructed as one of the background processes. This ignores the effects of
measurement errors, beam energy spread (partly due to bremsstrahlung), as well
as initial state radiation, but should nevertheless give a reasonable
indication of the effects of cuts that have to be imposed to isolate the
signal.

%%%%%%%%%%%%%%%%%%%%%%%%%%%%%%%%%%%%%%%%
\section{Effective asymmetries}
\label{sec:sec5}
\setcounter{footnote}{0}
%%%%%%%%%%%%%%%%%%%%%%%%%%%%%%%%%%%%%%%%

We are interested in constructing CP--odd observables. Schematically, they
are written as
\begin{eqnarray} \label{fdef}
F = \int d\Omega \frac{d\sigma}{d\Omega} f(\Omega) \times {\cal L} \, ,
\end{eqnarray}
where $d\sigma / d \Omega$ is the differential cross section, ${\cal L}=\int
L\, dt$ is the total integrated luminosity, and $f(\Omega)$ is a dimensionless
function of phase space observables. Introducing the luminosity in
Eq.~(\ref{fdef}) simplifies the statistical analysis as presented below.

Simple asymmetries are constructed from the choice $f = \pm 1$, where the
phase space region giving $f = +1$ is the CP--conjugate of that giving
$f=-1$~\cite{others,newbartl}. While very straightforward, this choice
usually does not yield the highest statistical significance. We decompose the
differential cross section into CP--even and CP--odd terms:
\begin{equation} \label{optcomp}
\frac {d \sigma} {d \Omega} = \sum_i e_i f^{(e)}_i(\Omega) +
\sum_j o_j f^{(o)}_j(\Omega)\, ,
\end{equation}
where the $e_i$ and $o_j$ are constant coefficients (products of
couplings and possibly masses) while the $f^{(e)}$ and $f^{(o)}$ are
CP--even and CP--odd functions, respectively, of phase space
variables. The optimal variable to extract the coefficient $o_j$ is
then proportional to $f^{(o)}_j$ \cite{optimal}.

In our case this would lead to very complicated observables, due
to the non--trivial angular dependence of the selectron propagators
$D_{(t,u)(L,R)}$ in Eq.~(\ref{bilinear}).
Moreover, the optimal
variables would depend on both selectron masses. For simplicity, we
construct our CP--odd observables by fully including the angular
dependence in the {\em numerators} of Eqs.~(\ref{diffx}),
(\ref{SUB}), (\ref{polcomp}), (\ref{SBU}), (\ref{SAB}) and
(\ref{decdist}), but ignoring the angular dependence in the
propagators.

For dimensionless $f$, the quantity $F$ in Eq.~(\ref{fdef}) is also
dimensionless. The statistical uncertainty of $F$ is then given by
\begin{eqnarray}
\sigma^2(F) = {\cal L}\times \int d\Omega \frac{d\sigma}{d\Omega}
f^2(\Omega) \, .
\end{eqnarray}
This can be seen from the fact that ${\cal L} (d \sigma / d \Omega) d \Omega$
is the number of events in the phase space interval $d \Omega$. For the simple
case of $f = \pm 1$, $\sigma^2(F)$ is simply the total number of events. With
the quantity $F$ and its statistical uncertainty $\sigma(F)$, we can construct
an effective asymmetry:
\begin{eqnarray} \label{ahat}
\hat{A}[f] = \frac{F}{\sigma(F) \sqrt{\cal L}}\, .
\end{eqnarray}
Note that $\hat A$ is by construction independent of the luminosity.  It is
also invariant under transformations $f(\Omega) \to c f(\Omega)$ for constant
$c$, making $\hat A$ independent of the normalization of $f$. The statistical
significance for $\hat A[f]$ is
simply given by $\hat A[f] \cdot \sqrt{{\cal L}}$.

%%%%%%%%%%%%%%%%%%%%%%%%%%%%%%%%%%%%%%
\section{Numerical analysis}
\label{sec:sec6}
\setcounter{footnote}{0}
%%%%%%%%%%%%%%%%%%%%%%%%%%%%%%%%%%%%%%

We are now ready to present some numerical results. We will first
briefly discuss the relevant quartic charges that encode CP
violation, before discussing ``production'' and ``decay''
asymmetries.

\subsection{Quartic charges}

Table~\ref{tab:quartic} shows that the four quartic charges $Q_4, \, Q_6, \,
Q_4'$ and $Q_6'$ are CP-odd.
Equation (\ref{SUB}) shows that $Q'_6$ is responsible for the production--level
asymmetry, which requires transverse beam polarization.\footnote{We note in
  passing that the corresponding asymmetry for chargino production vanishes
  \cite{bartl_char}: there is no equivalent of the $\tilde e_R$ exchange
  diagram, and the relevant $2 \times 2$ matrix diagonalizing the chargino
  mass matrix does not contain a reparametrization invariant phase.} The
remaining three CP--odd quartic charges can be probed only via the $\tilde
\chi_2^0$ polarization.  Equations (\ref{SBU}) and (\ref{SAB}) show that $Q_4$
contributes even for unpolarized $e^\pm$ beams, whereas $Q_4'$ ($Q_6$) only
contributes in the presence of longitudinal (transverse) beam polarization.

\begin{figure}[ht!]
\begin{center}
\includegraphics[height=7.8cm,width=7cm,angle=270]{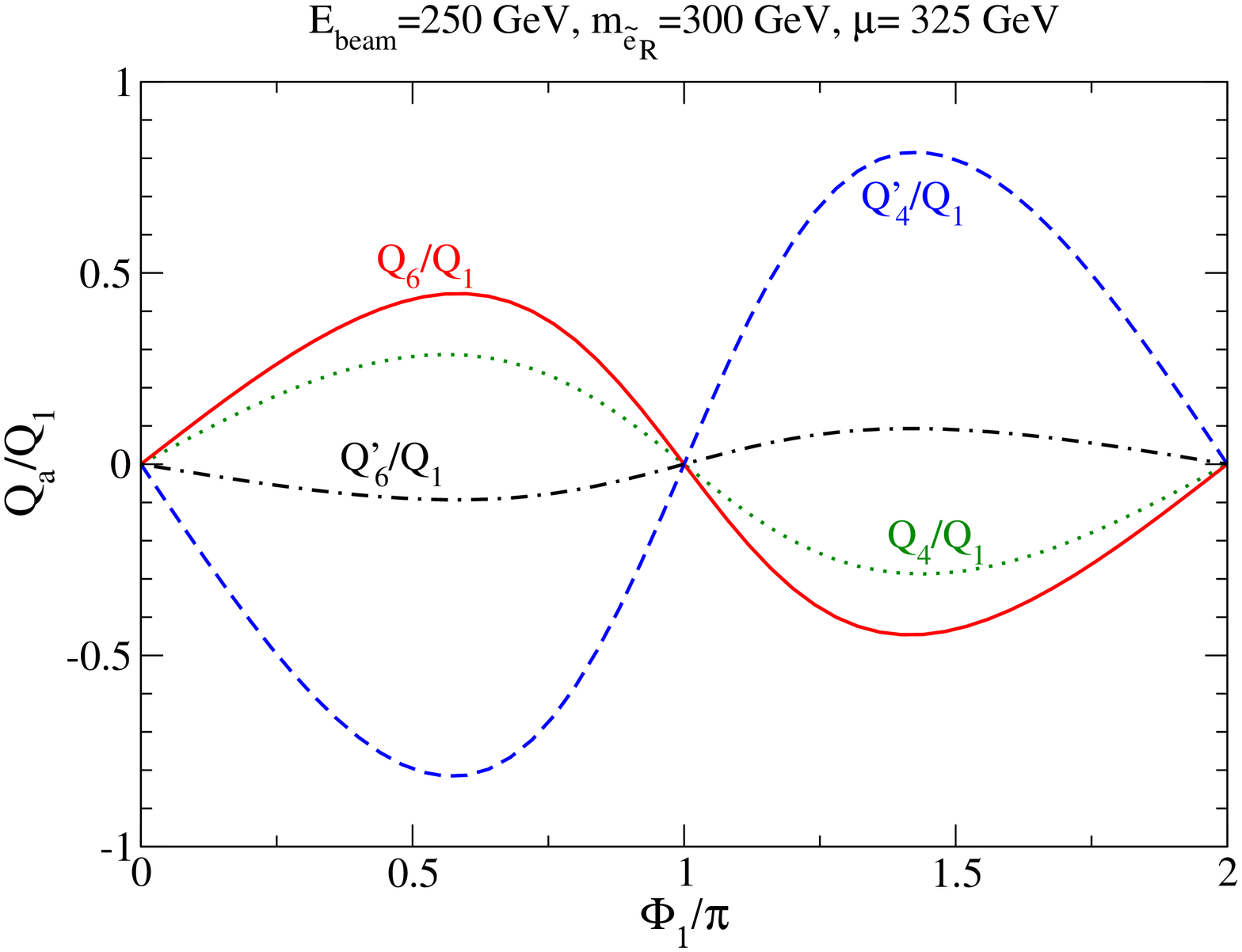} %\hskip 0.2cm
\includegraphics[height=7.8cm,width=7cm,angle=270]{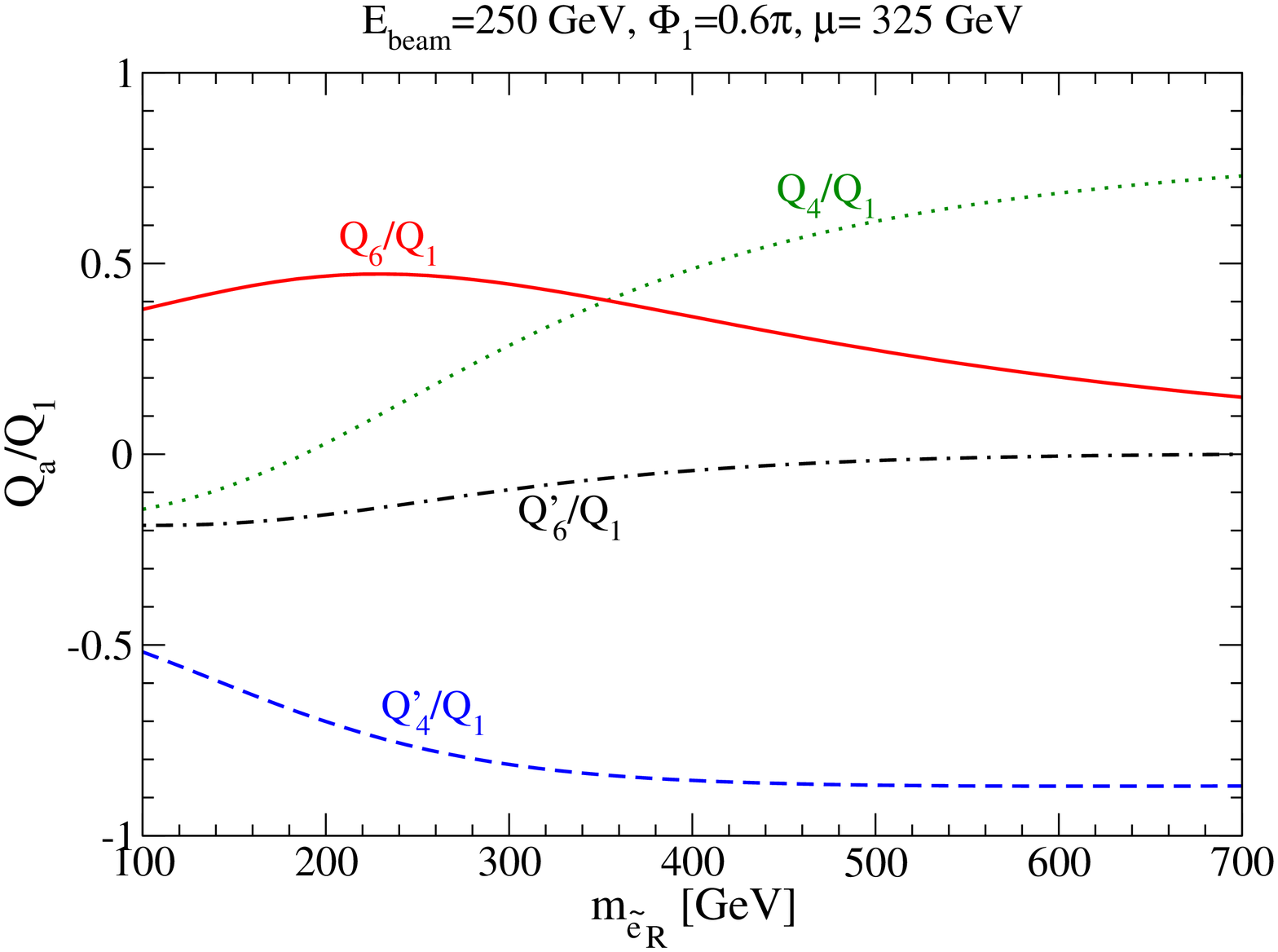}\\
\includegraphics[height=7.8cm,width=7cm,angle=270]{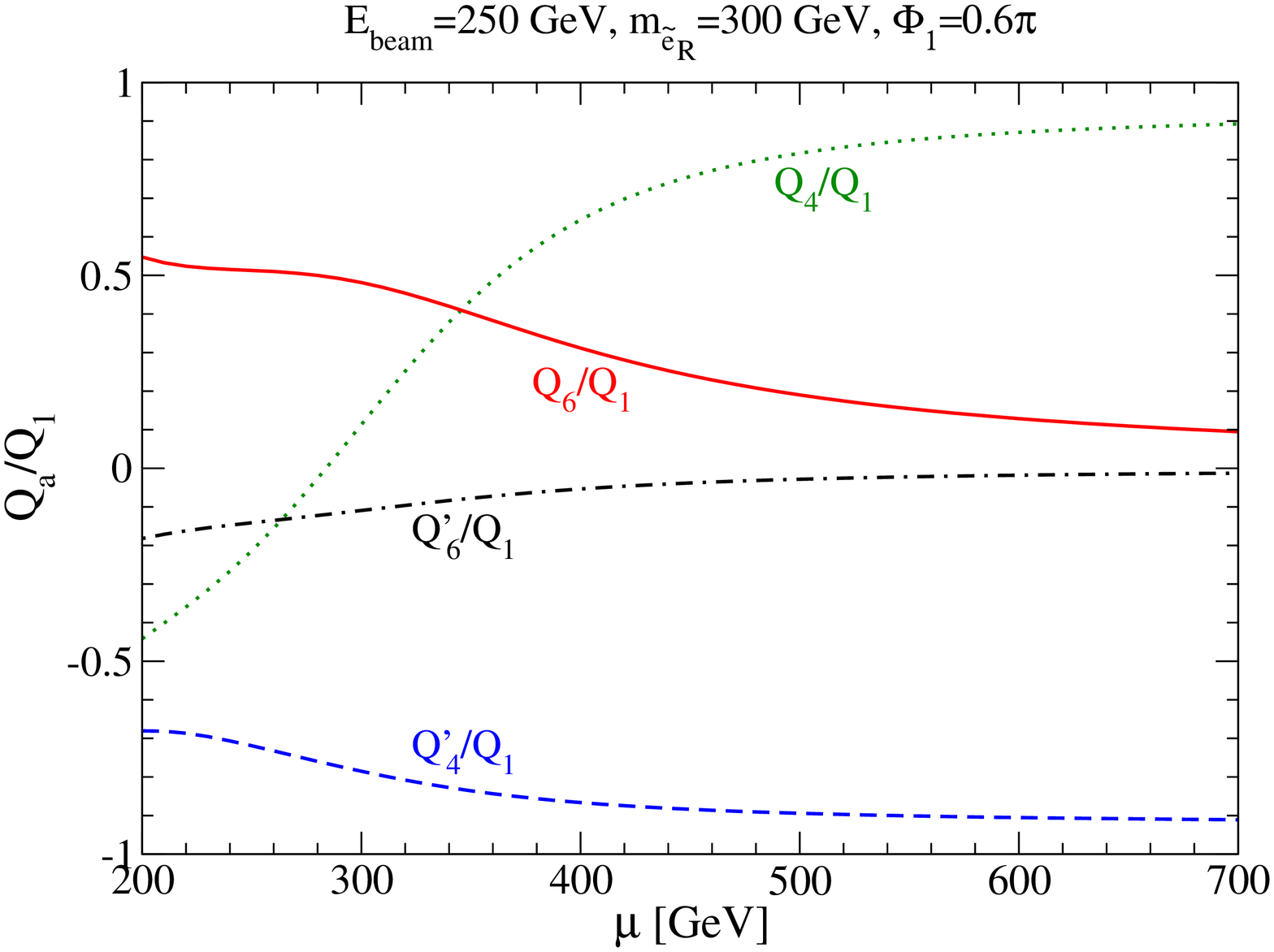} %\hskip 0.2cm
\includegraphics[height=7.8cm,width=7cm,angle=270]{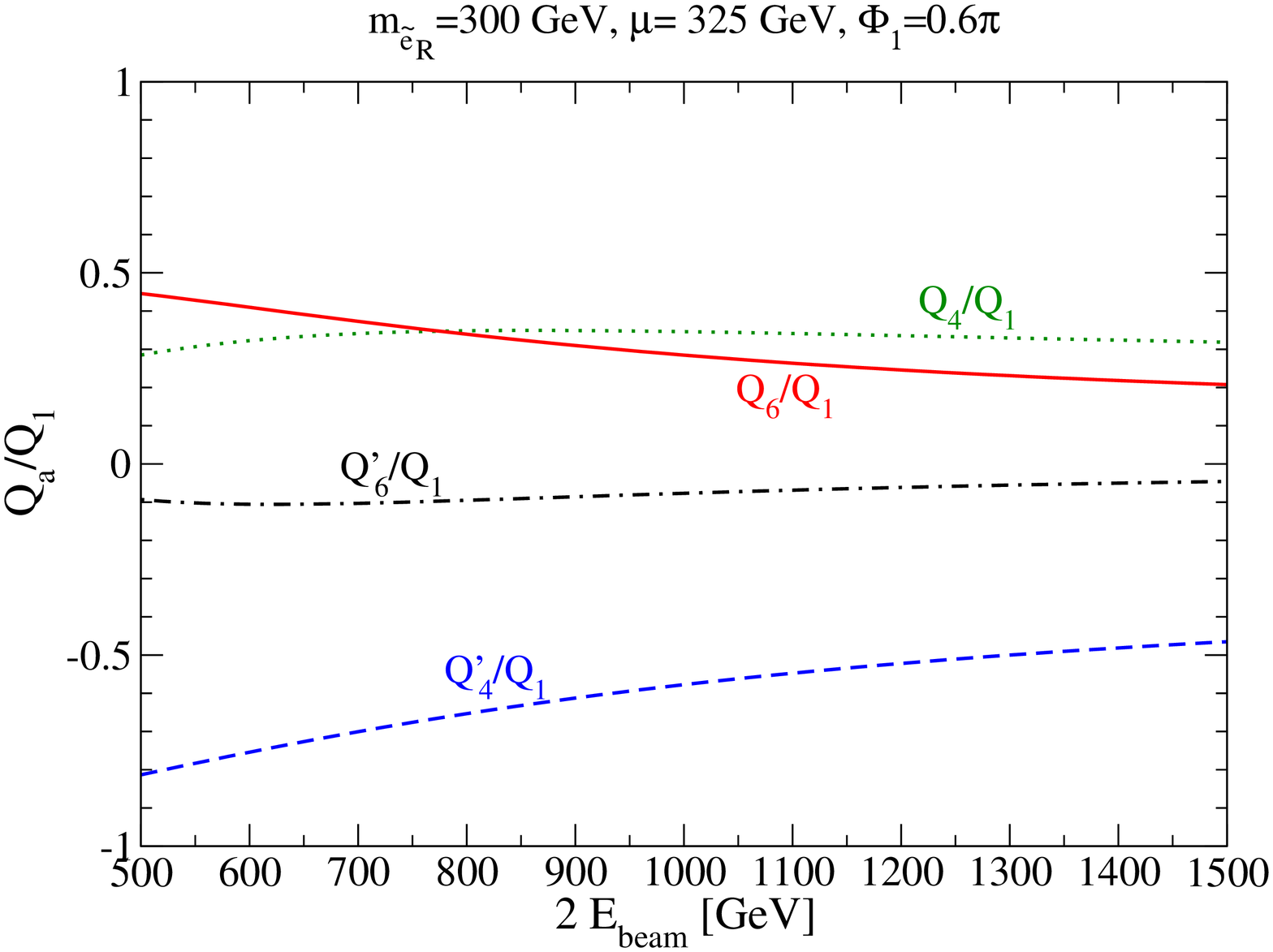}
\end{center}
\vskip -0.5cm
\caption{\it The ratios of quartic charges $Q_4/Q_1$ (dotted green),
  $Q'_4/Q_1$ (dashed blue), $Q_6/Q_1$ (solid red) and $Q'_6/Q_1$ (dot--dashed
  black). We fixed $|M_1| = 0.5 M_2 = 150$ GeV, $\tan\beta = 5, \ m_{\tilde
  e_L}= 500$ GeV and $\Phi_\mu = 0$; the values of the other relevant
  parameters are as indicated in the figures.}
\label{fig2}
\end{figure}

Figure~\ref{fig2} presents these four charges normalized to $Q_1$,
which largely determines the size of the unpolarized cross section
far above threshold. All these ratios lie between $-1$ and 1. We
took $|M_1| = 150$ GeV, $M_2 = 300$ GeV (so that $|M_1|$ and $M_2$
unify at the scale of Grand Unification \cite{book}), a moderate
$\tan\beta = 5$, $m_{\tilde e_L} = 500$ GeV, and $\Phi_\mu = 0$ (as
indicated by constraints on the electric dipole moments of the
electron and neutron \cite{edm,edm1}). The default choices of the
other relevant parameters are $|\mu| = 325$ GeV, $m_{\tilde e_R} =
300$ GeV, $\Phi_1 = 0.6 \pi$ and $\sqrt{s} = 2 E_{\rm beam} = 500$
GeV, but one of these parameters is varied in each of the four
frames of Fig.~\ref{fig2}. Finally, we chose scattering angle
$\cos\Theta = 1/\sqrt{2}\,$; note that $Q_6'$ vanishes at $\cos\Theta
= 0$.

The behavior of the curves in Fig.~\ref{fig2} can be understood with the help
of the expressions in Eq.~(\ref{quart}). The top--left frame shows the
dependence of the four ratios on the phase $\Phi_1$.  We see the typical
behavior of CP--odd quantities, changing sign when $\sin\Phi_1$ changes sign,
although not simple sine functions. Since we took $|\mu|$ to be close to
$M_2$, $\tilde \chi_2^0$ is a strongly mixed state. However, $\tilde \chi_1^0$
is still mostly gaugino--like, so that $|{\cal Z}_{12}|$ is quite small. As a
result, increasing $m_{\tilde e_R}$ (top--right frame) reduces $|Q_6|$ and
$|Q_6'|$, while affecting $|Q_4|$ and $|Q_4'|$ very little; recall that the
latter two quartic charges receive the dominant contribution from the
interference of $t-$ and $u-$channel $\tilde e_L$ exchange diagrams.
Increasing $|\mu|$ (bottom--left frame) has the same effect, as expected from
our earlier observation that $Q_6$ and $Q_6'$ need sizable gaugino--higgsino
mixing, while $Q_4$ and $Q_4'$ do not.  Finally, the bottom--right frame shows
that the dependence on the beam energy is relatively mild.

Another conclusion from Fig.~\ref{fig2} is that $|Q_6'|$ is usually the
smallest of the four CP--odd quartic charges. The reason is that in
this case $t-$ and $u-$channel diagrams tend to cancel, whereas they
add up in $|Q_6|$. This indicates that measuring the
production--level asymmetry will be quite challenging, as will be
discussed in the next Subsection.

\subsection{Production asymmetries}

The simplest choice for probing the CP--odd contribution from $ Q_6'$ to the
production cross section in Eq.~(\ref{diffx}) is \cite{newbartl}
\begin{equation} \label{fprod}
f_{\rm prod} = {\rm sign} [ \cos\Theta \sin(2\Phi)] \, .
\end{equation}
Instead a partly optimized asymmetry is suggested from the choice
\begin{equation} \label{fprodopt}
f_{\rm prod}^{\rm opt} = \cos\Theta \sin^2 \Theta \sin(2\Phi) \, ,
\end{equation}
where we have set the angle $\eta = 0$ for simplicity; nothing is
gained by considering nonvanishing angles between the transverse
$e^+$ and $e^-$ polarization vectors. The factors of $\sin^2 \Theta$
and $\sin(2\Phi)$ appear explicitly in the differential cross
section in Eq.~(\ref{diffx}); inclusion of the factor $\cos\Theta$,
which strictly speaking violates the construction principle
described in Sec.~5, is necessary in this case, since this
contribution to the cross section changes sign when $\cos\Theta \to
- \cos\Theta$.

Here it is appropriate to show that the asymmetries defined in
Eqs.~(\ref{fdef}), (\ref{fprod}) and (\ref{fprodopt}) are indeed CP--odd. This
can most easily be seen by using the so--called naive or $\widetilde {\rm T}$
transformation, which inverts the signs of all three--momenta and spins, but
(unlike a true ${\rm T}$--transformation) does not exchange initial and final
state. In the absence of absorptive phases\footnote{In the present context
  absorptive phases can only come from the finite width in the $Z-$propagator,
  which is entirely negligible for $s \gg m_Z^2$, or from loop corrections.} a
violation of $\widetilde{\rm T}$ invariance is equivalent to CP violation, as
long as CPT is conserved (which is certainly the case in the MSSM). Recall
that we fixed the $+z$ and $+x$ directions via the $e^-$ beam and spin
directions, respectively, which are themselves $\widetilde{\rm T}$ odd
quantities.\footnote{Note that for $\eta = 0$ the initial state is $\widetilde
  {\rm T}$ self--conjugate in this coordinate frame.} In this coordinate frame
a $\widetilde {\rm T}$ transformation therefore amounts to flipping the signs
of only the $y-$components of all three--momenta and spins. This is equivalent
to flipping the sign of the azimuthal angle $\Phi$ (as well as that of
$\Phi^*$, which is however irrelevant for the production--level asymmetry),
leaving $\Theta$ (and $\Theta^*$) unchanged. Our production--level asymmetries
are therefore $\widetilde {\rm T}$ odd, which probe
CP--violation if absorptive phases can be ignored.

\begin{figure}[ht!]
\begin{center}
\includegraphics[height=7.8cm,width=7cm,angle=270]{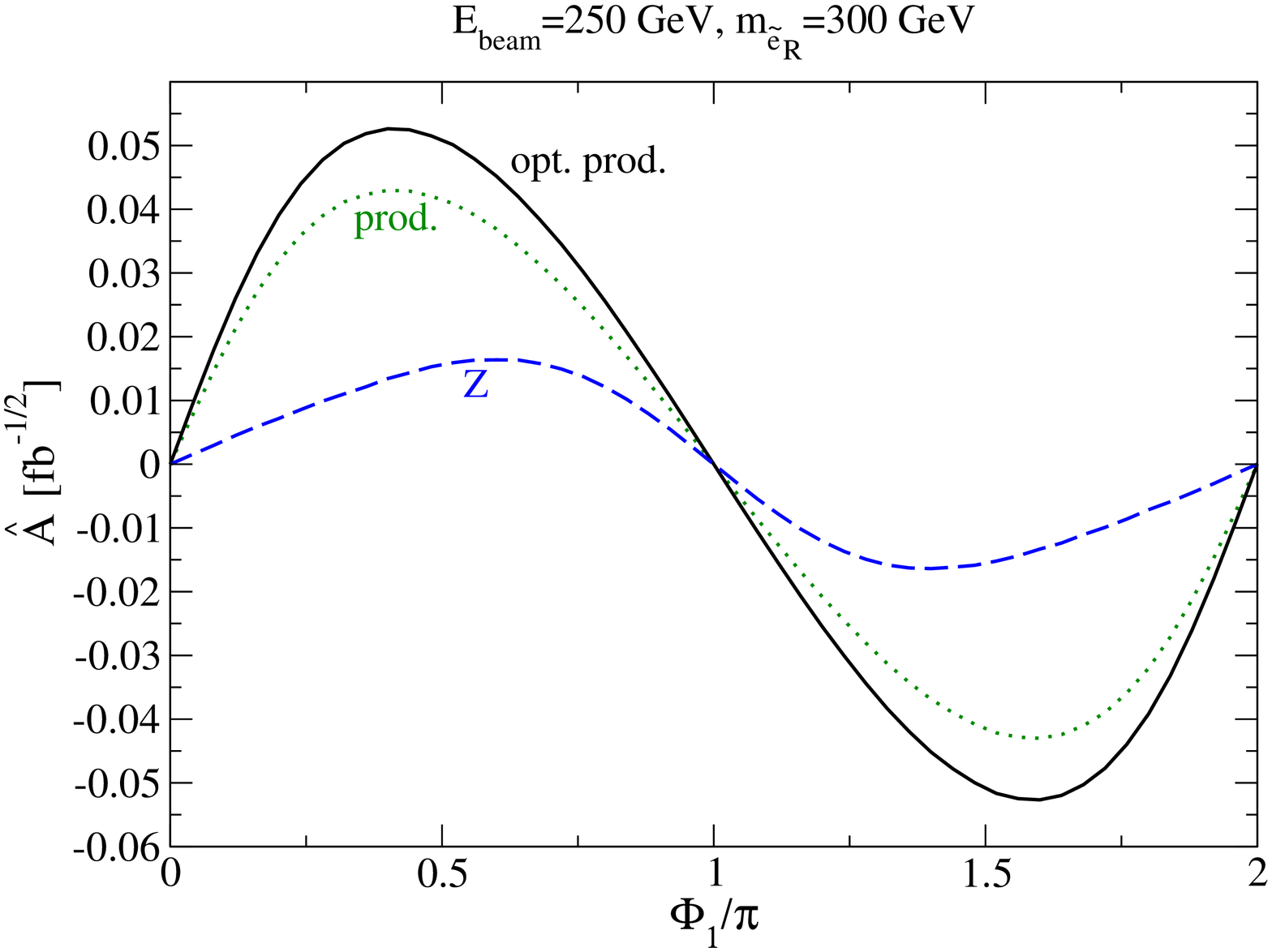} %\hskip 0.5cm
\includegraphics[height=7.8cm,width=7cm,angle=270]{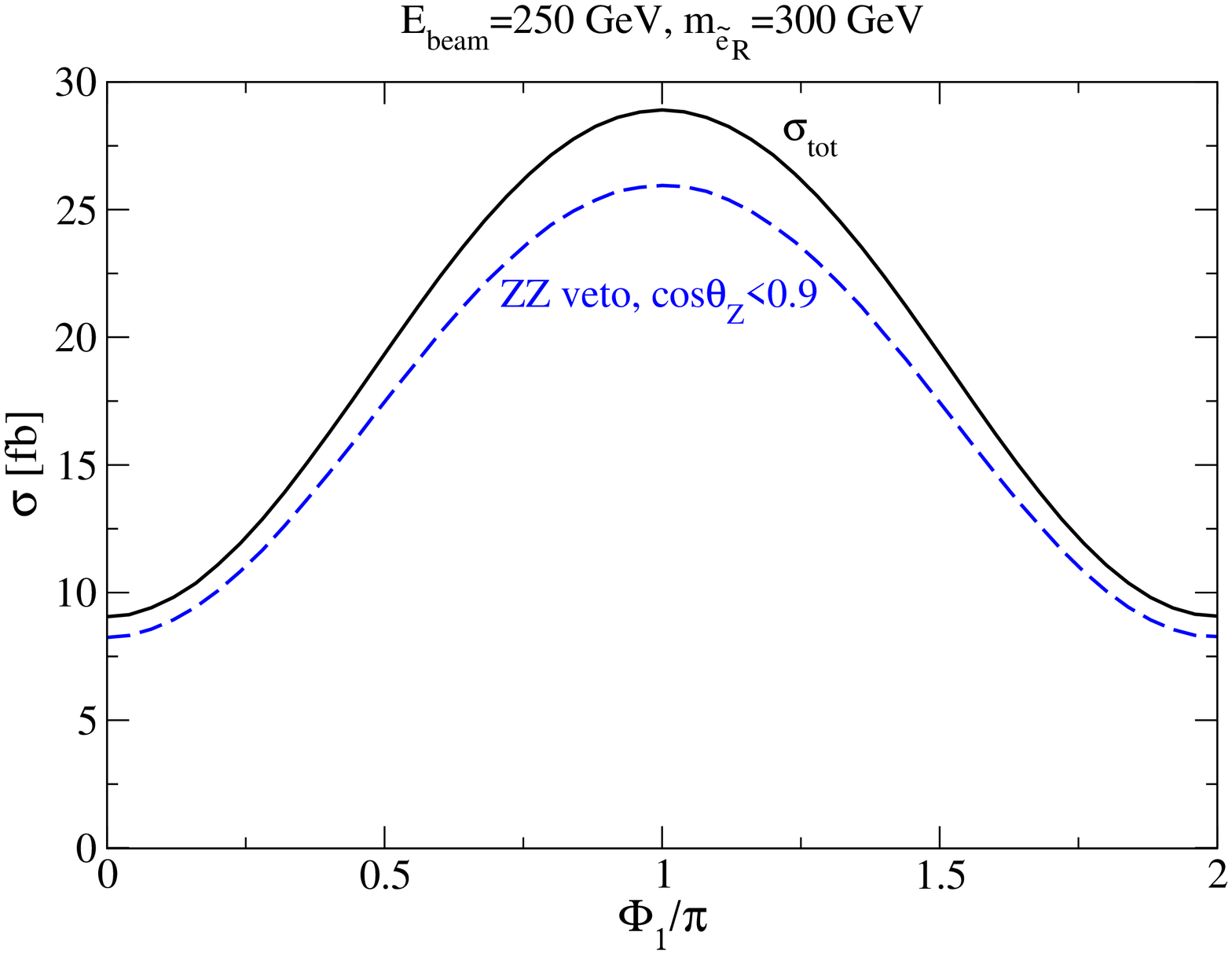}\\
\includegraphics[height=7.8cm,width=7cm,angle=270]{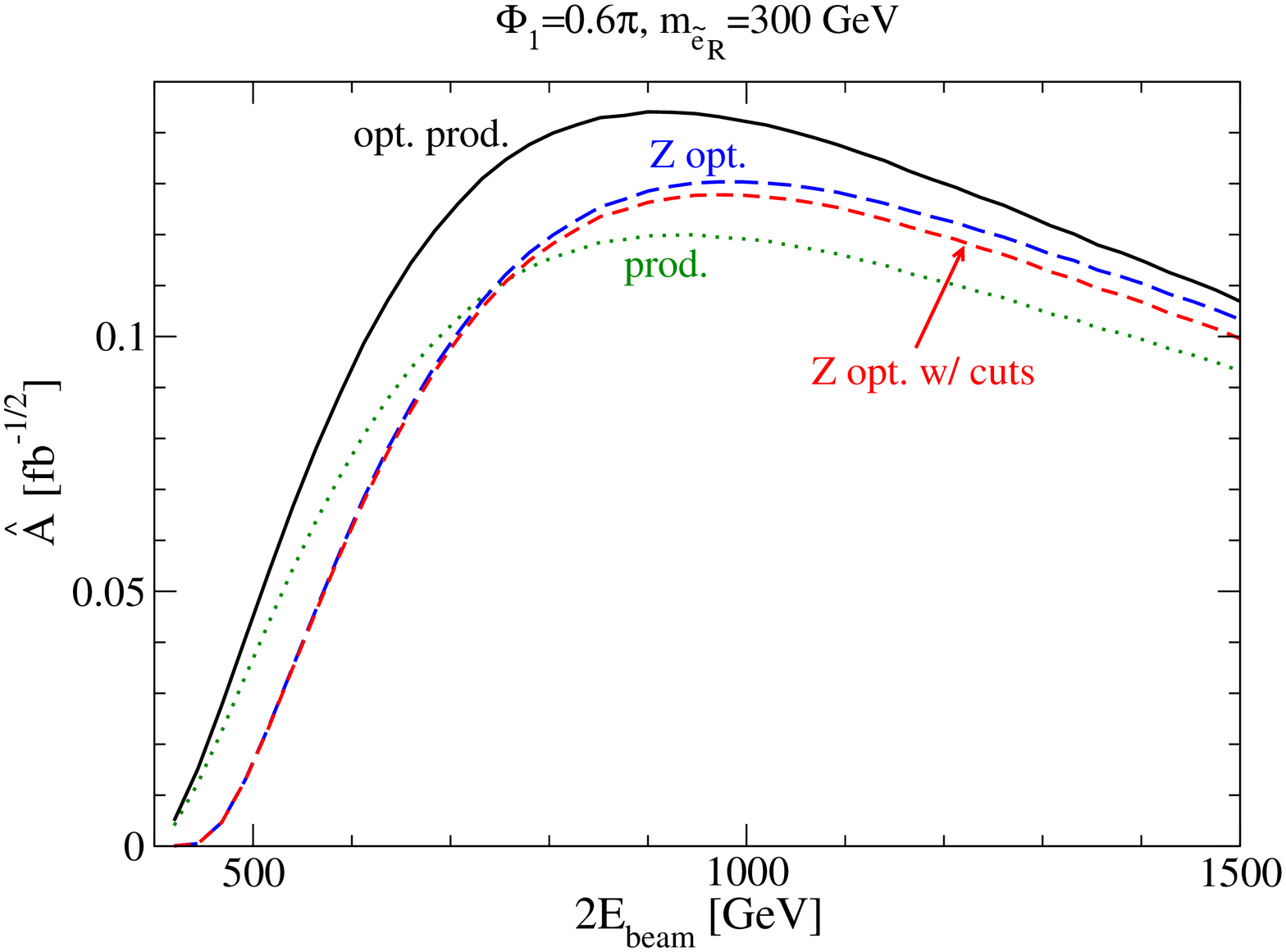} %\hskip 0.5cm
\includegraphics[height=7.8cm,width=7cm,angle=270]{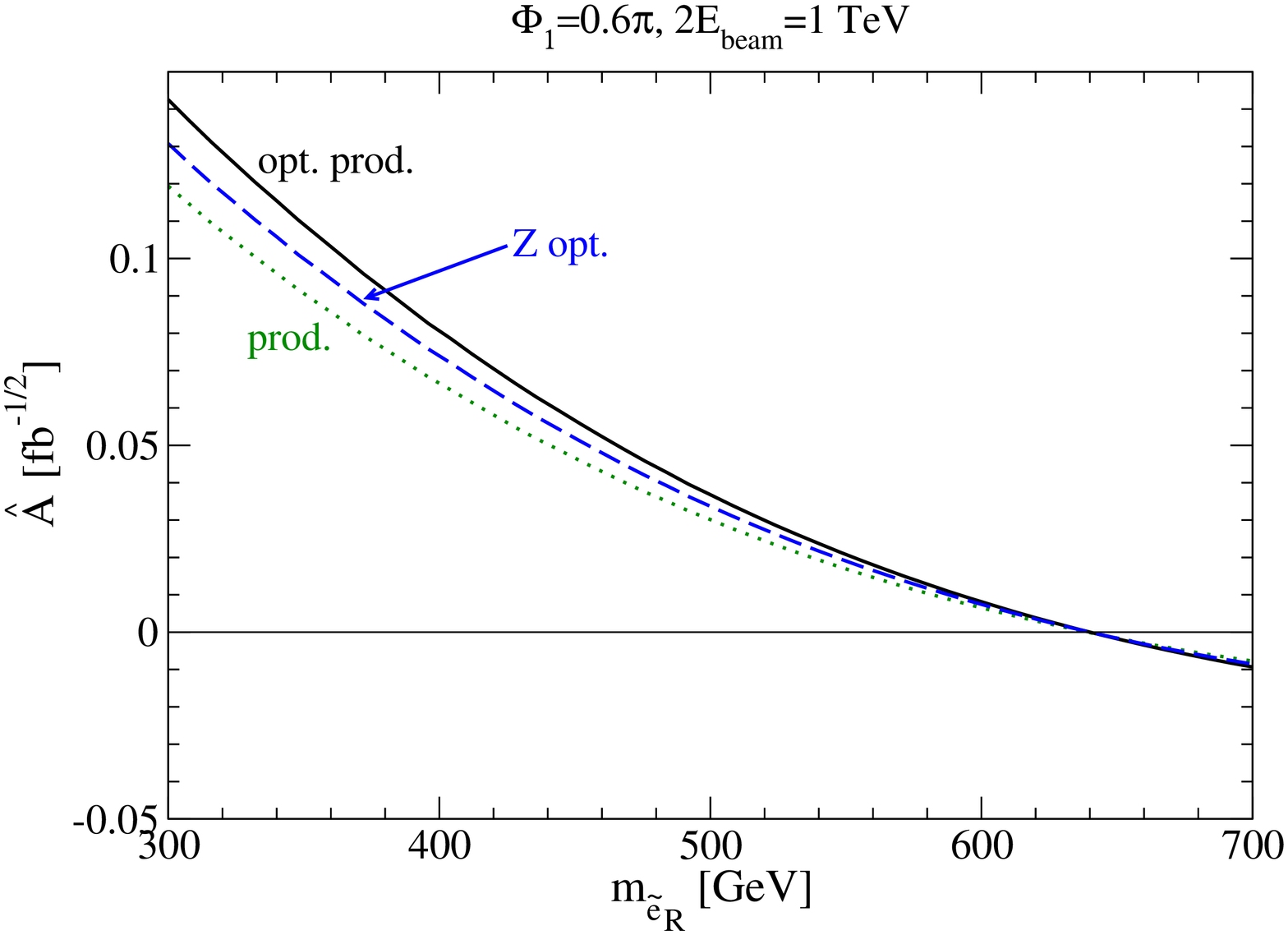}
\end{center}
\vskip -0.5cm
\caption{\it The top--left and both bottom frames show the effective
  production--level asymmetries defined by Eq.\,(\ref{fprod}) (green dotted
  curves, labeled ``prod.'') and (\ref{fprodopt}) (solid black curves, labeled
  ``opt. prod.''), together with the ``optimized'' production asymmetry where
  the true production angles are replaced by those reconstructed from the $Z$
  direction (blue long--dashed curves: without cuts; red short--dashed curve:
  with the cuts described in the text). The top--right frame shows the total
  cross section for $e^+e^- \to \tilde \chi_1^0 \tilde \chi_2^0$ without
  (black solid curve) and with (blue dashed curve) cuts. The default
  parameters are as in Fig.~\ref{fig2}, but one parameter is varied in each
  frame.}
\label{fig3}
\end{figure}

The effective asymmetries resulting from Eqs.~(\ref{fprod}) and
(\ref{fprodopt}) are shown by the (green) dotted and (black) solid curves,
respectively, in three frames in Fig.~\ref{fig3}. In these figures we have
chosen the same default parameters as in Fig.~\ref{fig2}, which ensures that
$\tilde \chi_2^0 \to \tilde \chi_1^0 Z$ is the only possible two--body decay
of $\tilde \chi_2^0$.\footnote{The effective asymmetry constructed from
  $\tilde \chi_2^0 \to \tilde \chi_1^0 h$ decays is very similar to that from
  $\tilde \chi_2^0 \to \tilde \chi_1^0 Z$ decays; we therefore do not show
  numerical results for this decay mode.} As noted in Sec.~3, in this case we
can measure the $\tilde \chi_2^0$ polarization only if the polarization of the
$Z$ boson is determined. In particular, one has to be able
to distinguish between the two transverse polarization states in order to
construct CP--odd asymmetries involving the $Z$ polarization. Although this
measurement is, in principle, possible for $Z \rightarrow \ell^+ \ell^-$
decays, the efficiency is quite low due to its small branching ratio ($\sim
7\%$ after summing over $e$ and $\mu$ final states), and a very poor analyzing
power (from almost purely axial vector coupling for $Z\ell^+ \ell^-$).
Although $q
\bar q$ final states have larger analyzing power,
the measurement of
the charge is very difficult. It may be only possible to probe the
production level asymmetry through this decay mode.

Unfortunately the event cannot be reconstructed in this mode, as
noted in Sec.~4. This means that we do not know the angles $\Theta$
and $\Phi$ appearing in the definitions of Eqs.~(\ref{fprod}) and
(\ref{fprodopt}); the best we can do is to approximate them by the
corresponding angles of the $Z$ boson. This leads to the (blue)
dashed curves in the frames of Fig.~\ref{fig3} that show effective
asymmetries, which are based on the ``optimized'' choice in
Eq.~(\ref{fprodopt}).

The top--left frame shows these asymmetries as functions of the
CP--odd phase $\Phi_1$. We see that the ``optimized'' effective
asymmetry exceeds the simple asymmetry based on Eq.~(\ref{fprod}) by
typically $\sim 20\%$, leading to a $\sim 40\%$ reduction of the
luminosity required to establish the existence of a non--vanishing
asymmetry at a given confidence level. Unfortunately replacing the
true production angles ($\Theta$ and $\Phi$) by those of the $Z$ boson
reduces the effective asymmetry by a factor of 2.5$-$3.5.
This suppression factor depends on the masses of the two lightest
neutralinos, which in turn depend on $\Phi_1$. In this case even for
the most favorable choice of parameters an integrated luminosity of
several ab$^{-1}$ would be needed to establish a non--vanishing
optimized asymmetry at the $1\sigma$ level, even assuming 100\% beam
polarization! This is well beyond the currently expected
performance of the international linear collider.

The lower--left frame of Fig.~\ref{fig3} shows that the situation
might be better at higher beam energies. The effective production
asymmetries peak at $\sqrt{s} \simeq 900$ GeV for the given choice
of SUSY parameters. Moreover, the difference between the
``theoretical'' optimized asymmetry and the one constructed from the
$Z$ boson angles becomes much smaller at higher energy. The reason
is that at $\sqrt{s} \gg m_{\tilde \chi_2^0}$ the $\tilde{\chi}^0_2$
becomes ultra--relativistic; its decay products then fall in a
narrow cone around the $\tilde \chi_2^0$ direction, so that the
differences between the real production angles ($\Theta$ and $\Phi$)
and the corresponding angles derived from the flight direction of
the $Z$ boson become small. However, even in this case 1 ab$^{-1}$
would only allow to establish an asymmetry with a significance of
3.5 standard deviations at best, ignoring experimental resolutions
and efficiencies, and assuming 100\% transverse beam polarization.
The bottom--right frame shows that the situation is even worse if
the mass of the SU(2) singlet selectron $\tilde e_R$ is close to
that of the SU(2) doublet $\tilde e_L$, which is taken as 500 GeV in
this figure.

The top--right figure is a reminder that $\tilde \chi_1^0 \tilde \chi_2^0$
production can nevertheless  provide useful information on the phase $\Phi_1$
\cite{Choi:2004rf}, simply through a measurement of the total production cross
section, which increases by almost a factor of three when $\Phi_1$ is varied
from $0$ to $\pi$; no beam polarization is needed for this measurement. As
explained in Refs.~\cite{Choi:2001ww,Choi:2004rf} this is due to the fact that
the production occurs in a pure $P-$wave for $\Phi_1=0$, but has a large
$S-$wave component for $\Phi_1 = \pi$. This figure also shows that, for the
chosen set of parameters, cutting against the $ZZ$ background as described in
Sec.~\ref{sec:sec4}, as well as applying the acceptance cut
\begin{equation} \label{cut}
|\cos\Theta_X| \leq 0.9
\end{equation}
for all visible final state particles $X$ (in this case, the $Z$
boson), only reduces the cross section by $\sim15\%$. The (red)
short--dashed curve in the bottom--left frame shows that these cuts
affect the effective asymmetries even less.

\subsection{Decay asymmetries}
\setcounter{footnote}{0}

We now turn to the ``decay'' asymmetries, which are
sensitive to the $\tilde \chi_2^0$ polarization. We saw in
Sec.~\ref{sec:sec3} that these can be only probed through $\tilde
\chi_2^0 \to \tilde \ell^\pm \ell^\mp$ decays (ignoring three--body
decays, which will be highly suppressed if any two--body decay is
allowed). The discussion of Sec.~\ref{sec:sec4} showed that in this
case we can reconstruct the event with two-- or four--fold
ambiguity.

Equation~(\ref{polcomp}) shows that there are three CP--odd terms in the
$\tilde \chi_2^0$ polarization vector, which are sensitive to transverse beam
polarization. In order to construct the corresponding ``optimized''
asymmetries, we first need an explicit expression for the scalar product
appearing in Eq.~(\ref{decdist}).  Working in the reference frame where the
$+x$ direction is defined by the transverse part of the $e^-$ polarization
vector, and using the same set of axes for the definition of the $\tilde
\chi_2^0$ decay angles $\Theta^*, \Phi^*$ in the $\tilde \chi_2^0$ rest frame,
we find using Eqs.~(\ref{polvec1}) and (\ref{polvec2}):
\begin{eqnarray} \label{pk}
\overrightarrow{{\cal P}} \cdot \hat{ k}_1^* &=&\,\,
{\cal P}_T \left[ \cos \Theta \sin \Theta^*  \cos (\Phi - \Phi^*) - \sin\Theta
  \sin \Theta^* \right] \nonumber  \\
&&\!\!\!+\, {\cal P}_L \left[ \sin\Theta \sin\Theta^* \cos(\Phi - \Phi^*)
+ \cos  \Theta \cos \Theta^* \right] \nonumber \\
&&\!\!\!+\, {\cal P}_N \sin \Theta^* \sin(\Phi - \Phi^*) \, ,
\end{eqnarray}
where we have suppressed the superscript 2 on the components of the $\tilde
\chi_2^0$ polarization vector. This, together with Eqs.~(\ref{polcomp}) and
(\ref{SAB}), leads to the following choices for $f$ in
Eq.~(\ref{fdef}):\footnote{Note that the denominator $\Delta_U^{21}$ in
  Eq.~(\ref{pdel}) cancels against the factor $\Delta_U^{21}$ from the
  production cross section (\ref{diffx}) in the final result for the cross
  section differential in production and decay angles.}
\begin{eqnarray} \label{ftopt}
f_{LN} &=& \left[ \sin\Theta \sin\Theta^* \cos(\Phi - \Phi^*) + \cos
  \Theta \cos \Theta^* \right] \sin(2\Phi) \sin^2 \Theta\, , \nonumber \\
f_{TN} &=& \left[ \cos \Theta \sin \Theta^*  \cos (\Phi - \Phi^*) - \sin\Theta
  \sin \Theta^* \right] \sin(2\Phi) \sin(2\Theta) \, , \nonumber \\
f_{NT} &=& \left[ \sin \Theta^* \sin(\Phi - \Phi^*) \right] \cos(2\Phi) \sin
\Theta \, .
\end{eqnarray}
In each of the three expressions the factor in square brackets comes
from Eq.~(\ref{pk}), the second factor from Eq.~(\ref{polcomp}), and
the last factor from the expressions for $\Sigma_{LN}, \,
\Sigma_{TN}$ and $\Sigma_{NT}$, respectively, in Eq.~(\ref{SAB}).

Similarly, the expression for $\Delta_N^{21}$ in Eq.~(\ref{polcomp}) contains
two CP--odd terms that can be probed with only longitudinal beam polarization,
or even with unpolarized beams. Since the expressions for $\Sigma_{NU}$ and
$\Sigma_{NL}$ in Eqs.~(\ref{SBU}) and (\ref{SAB}) are identical except
for different quartic charges, we can combine these two terms into
the ``optimized'' longitudinal effective asymmetry $\hat{A}_L \equiv
\hat{A}[f_L]$ with
\begin{equation} \label{fl}
f_L = \left[ \sin \Theta^* \sin(\Phi - \Phi^*) \right] \sin \Theta\, .
\end{equation}
Note that the four functions $f_i$ defined in Eqs.~(\ref{ftopt}) and
(\ref{fl}) are all orthogonal to each other, i.e., the product of any
two different functions will vanish when integrated over the entire
phase space.

Although the three asymmetries defined in Eqs.~(\ref{ftopt}) are independent
of each other (probing different $\Sigma_{AB}$), in the context of the MSSM
they all probe the same quartic charge $Q_6$. If $m_{\tilde \chi_1^0}$ and
$m_{\tilde \chi_2^0}$ are known, one can therefore construct a single
asymmetry to probe $Q_6$, called the total ``optimized" transverse decay
asymmetry $\hat{A}_T \equiv \hat{A}[f_T]$ with
\begin{eqnarray} \label{ft}
f_T &=&  \left[ \sin\Theta \sin\Theta^* \cos(\Phi - \Phi^*) + \cos
  \Theta \cos \Theta^* \right] \sin(2\Phi) \sin^2 \Theta \cdot \left( 1 +
  \mu_1^2 - \mu_2^2 \right)
\nonumber \\
&+&  \left[ \cos \Theta \sin \Theta^*  \cos (\Phi - \Phi^*) - \sin\Theta
  \sin \Theta^* \right] \sin(2\Phi) \sin(2\Theta) \cdot \mu_2
\nonumber \\
&+&  \left[ \sin \Theta^* \sin(\Phi - \Phi^*) \right] \cos(2\Phi) \sin
\Theta \cdot 2 \mu_2 \, ,
\end{eqnarray}
where the $\mu_i$ have been defined in Eq.~(\ref{mu}).  The first,
second and third line in Eq.~(\ref{ft}) correspond to the
contributions from $\Sigma_{LN}, \, \Sigma_{TN}$ and $\Sigma_{NT}$,
respectively.

Finally, we also consider an effective asymmetry based on the
measurement of the momentum of the positive lepton $\ell_1$ coming from the
first stage of $\tilde \chi_2^0$ decay, defined by $\hat{A}_1^+
\equiv\hat{A}[f_1^+]$ with
\begin{equation} \label{f1}
f_1^+ = \sin(2  \Phi_{\ell_1^+})\, .
\end{equation}
The advantage of this asymmetry, which is somewhat similar to the
decay asymmetry considered in Ref.~\cite{newbartl}, is that it does
not need event reconstruction, as long as the ``primary'' and
``secondary'' leptons can be distinguished.

As discussed in the previous Subsection, a CP--odd observable changes sign
when $\Phi \to - \Phi$ and $\Phi^* \to - \Phi^*$.  Evidently the asymmetries
defined in Eqs.~(\ref{ftopt}) through (\ref{f1}) satisfy this condition. Due
to the sign flip in Eq.~(\ref{decdist}) all asymmetries discussed in this
Subsection have opposite signs for $\tilde \chi_2^0 \to \tilde \ell_R^+
\ell^-$ and $\tilde \chi_2^0 \to \tilde \ell_R^- \ell^+$ decays; events of
these two kinds should be treated separately. Since there are equal number of
events from these two decay chains, we can simply focus on events with only
positively charged primary leptons.

\begin{figure}[ht!]
\begin{center}
\includegraphics[height=7.8cm,width=7cm,angle=270]{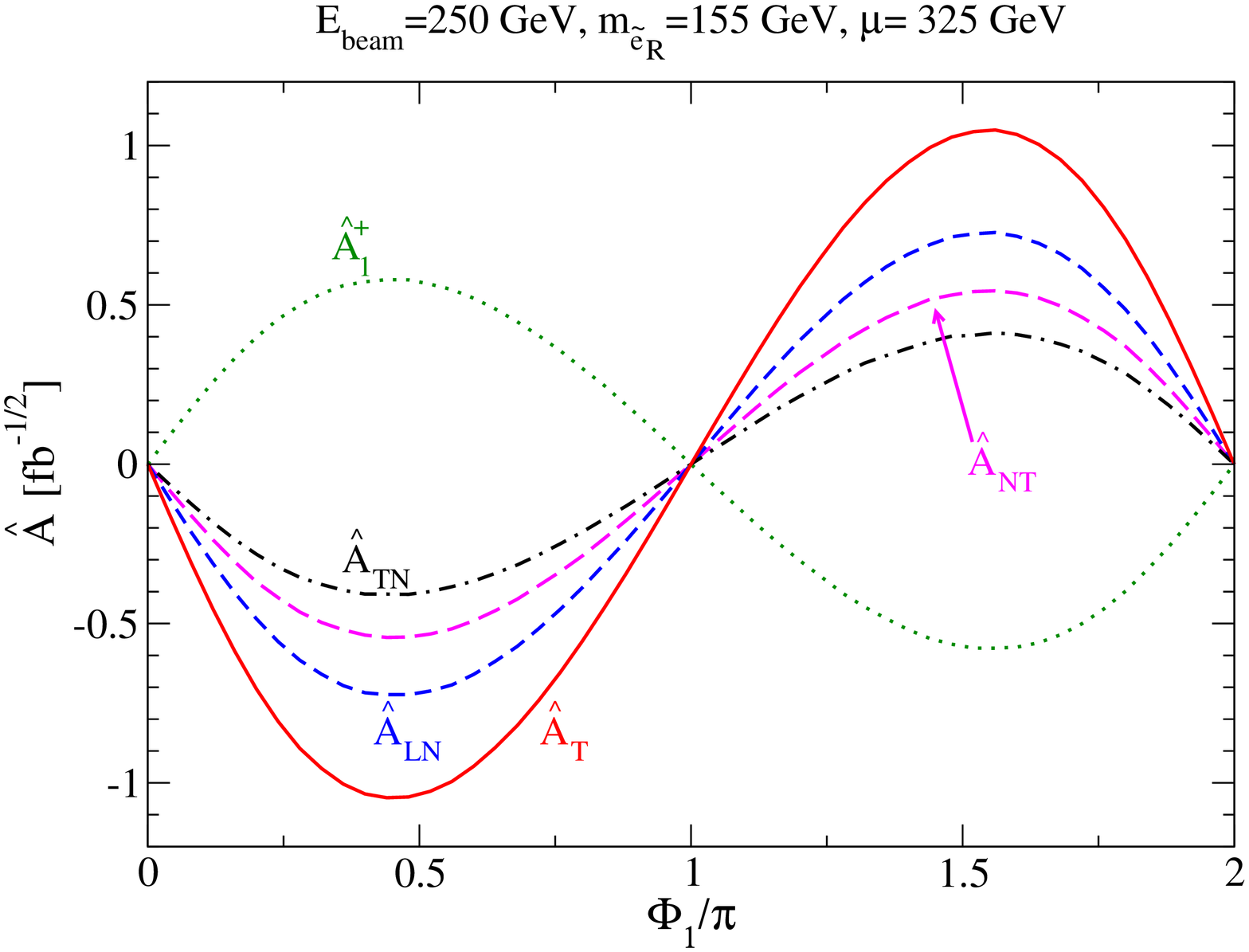} %\hskip 0.5cm
\includegraphics[height=7.8cm,width=7cm,angle=270]{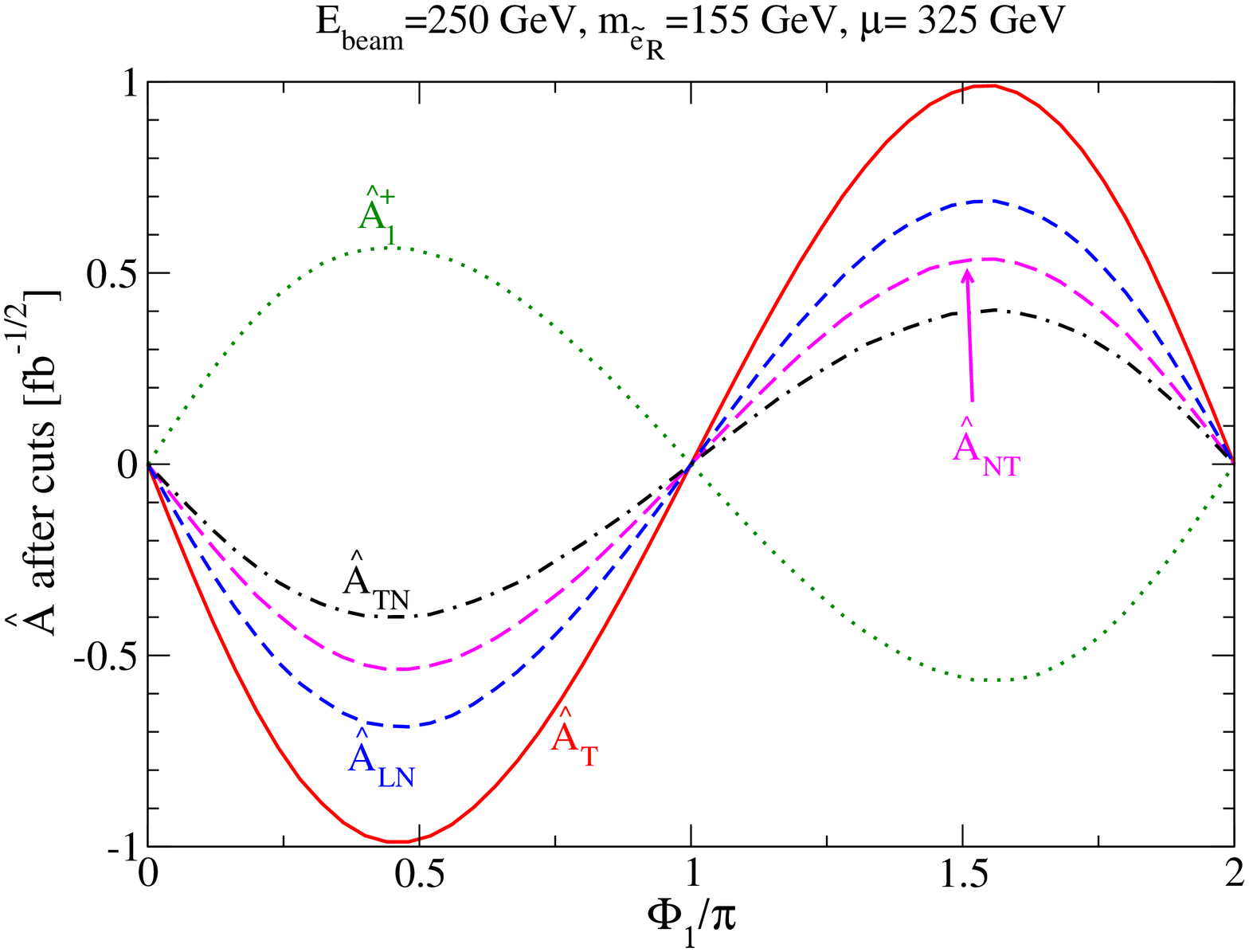}
\end{center}
\vskip -0.5cm
\caption{\it Effective transverse decay asymmetries for the same default
  parameters as in Fig.~\ref{fig2}, except that now $m_{\tilde e_R} = 155$
  GeV.  The (black) dot--dashed, (magenta) long dashed and (blue) short dashed
  curves show the ``optimized'' asymmetries based on $f_{TN}, \, f_{NT}$ and
  $f_{LN}$ in Eq.~(\ref{ftopt}), respectively, while the (red) solid curves
  show $\hat{A}_T$ of Eq.~(\ref{ft}), and the (green) dotted curves show
  $\hat{A}^+_1$ of Eq.~(\ref{f1}). In the right (left) frame acceptance and
  background--removing cuts have (not) been applied.}
\label{fig4}
\end{figure}

The two figures in Fig.~\ref{fig4} show the effective ``optimized'' decay
asymmetries based on Eqs.~(\ref{ftopt}), (\ref{ft}) and (\ref{f1}). We use the
same default parameters as in Figs.~\ref{fig2} and \ref{fig3}, except that the
$\tilde e_R$ mass has been reduced to 155 GeV, so that $\tilde \chi_2^0 \to
\tilde e_R^\pm e^\mp$ decays are allowed and dominant.
% (together with other
%leptonic decays, if sleptons are degenerate in mass).
Our choice of $m_{\tilde e_R}$ implies that $m_{\tilde \chi_2^0} - m_{\tilde
  e_R} \gg m_{\tilde e_R} - m_{\tilde \chi_1^0}$. As discussed in
Sec.~\ref{sec:sec4} this implies that the harder lepton always comes from the
first step of $\tilde \chi_2^0$ decay, allowing to reconstruct the event with
only a two--fold ambiguity. We average over both of these solutions when
calculating the ``optimized'' asymmetries. We find that the wrong
reconstruction typically leads to asymmetries with the same sign as the true
solution, with (of course) smaller magnitude. The dilution of the asymmetries
due to the event reconstruction ambiguity is therefore not very severe. The
effective asymmetry based on $f_{LN}$ of Eq.~(\ref{ftopt}) and, especially,
the one based on $f_T$ of Eq.~(\ref{ft})  are therefore substantially
larger in magnitude than the simple effective asymmetry based on Eq.~(\ref{f1}).
Note also that the three effective asymmetries based on Eq.~(\ref{ftopt})
move ``in step'', as expected from our earlier observation that they all probe
the same quartic charge $Q_6$. Combining them into a single effective asymmetry,
as in Eq.~(\ref{ft}), therefore increases the size of the asymmetry significantly.

The two frames in Fig.~\ref{fig4} differ in that the left figure does not
include any cuts whereas in the right figure we remove events that can be
reconstructed as $W$ or $\tilde{e}_R$  pair background events. Also, we
apply the acceptance cut in Eq.~(\ref{cut}) to both final state leptons.
For the case at hand these cuts only reduce the effective asymmetries by
10\% to 20\%. This high cut efficiency is also due to our choice of
masses, which implies that the two leptons in the final state have
very different energies. In contrast, both background processes have
identical energy distributions for the two
leptons in the final state. Signal events can be rarely reconstructed as
background in this scenario. As a result we find that even after cuts one
would only need an integrated luminosity of $\sim 10$ fb$^{-1}$ to measure a
non--vanishing asymmetry at the $3\sigma$ level. This still assumes 100\% beam
polarization. Even for the more realistic choice $P_T \overline{P}_T
\simeq 0.5$ one might achieve $3\sigma$ significance with $\sim 40$ fb$^{-1}$
of data.  This integrated luminosity should be achievable, assuming that
transverse beams will be available.

\begin{figure}[ht!]
\begin{center}
\includegraphics[height=7.8cm,width=7cm,angle=270]{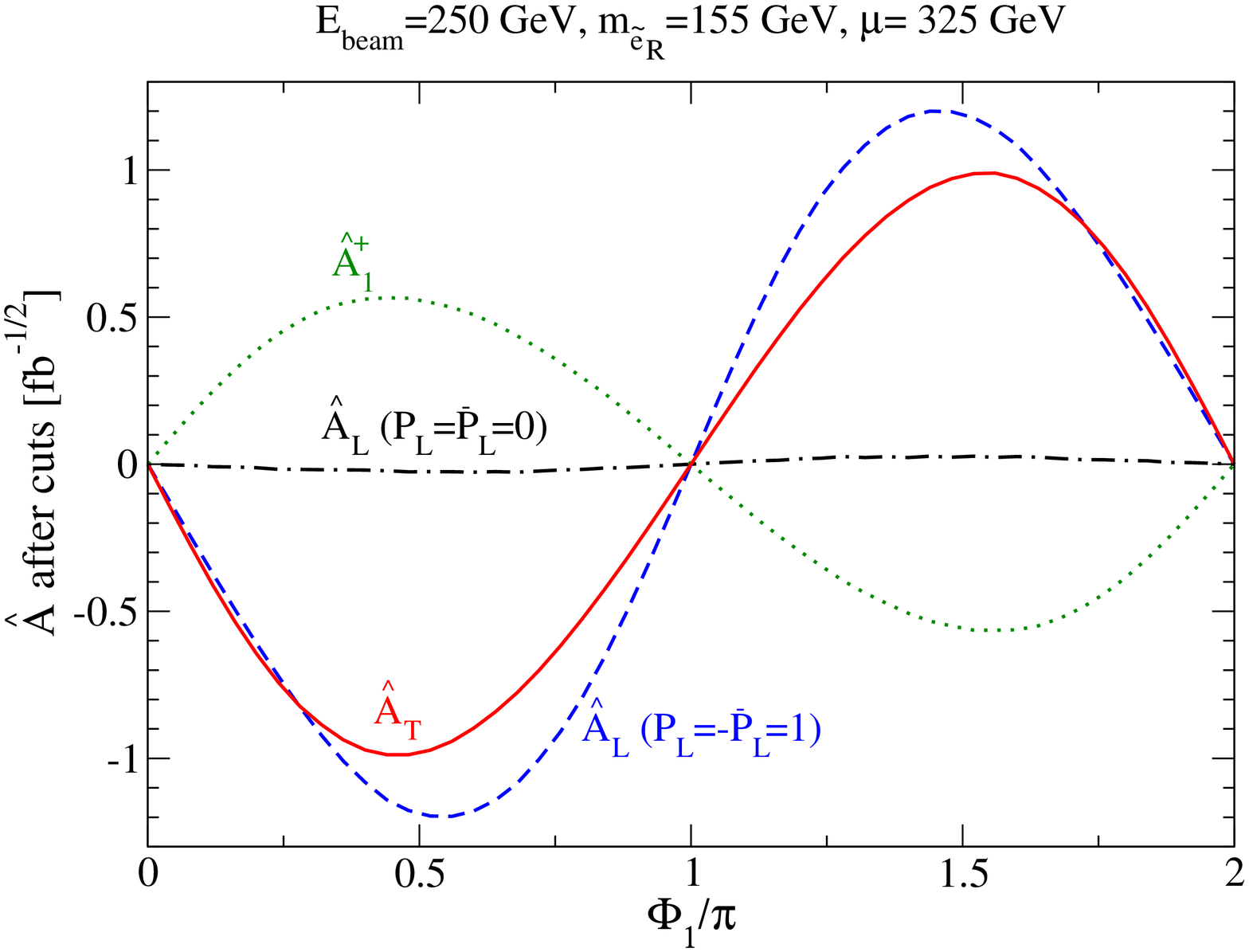} %\hskip 0.5cm
\includegraphics[height=7.8cm,width=7cm,angle=270]{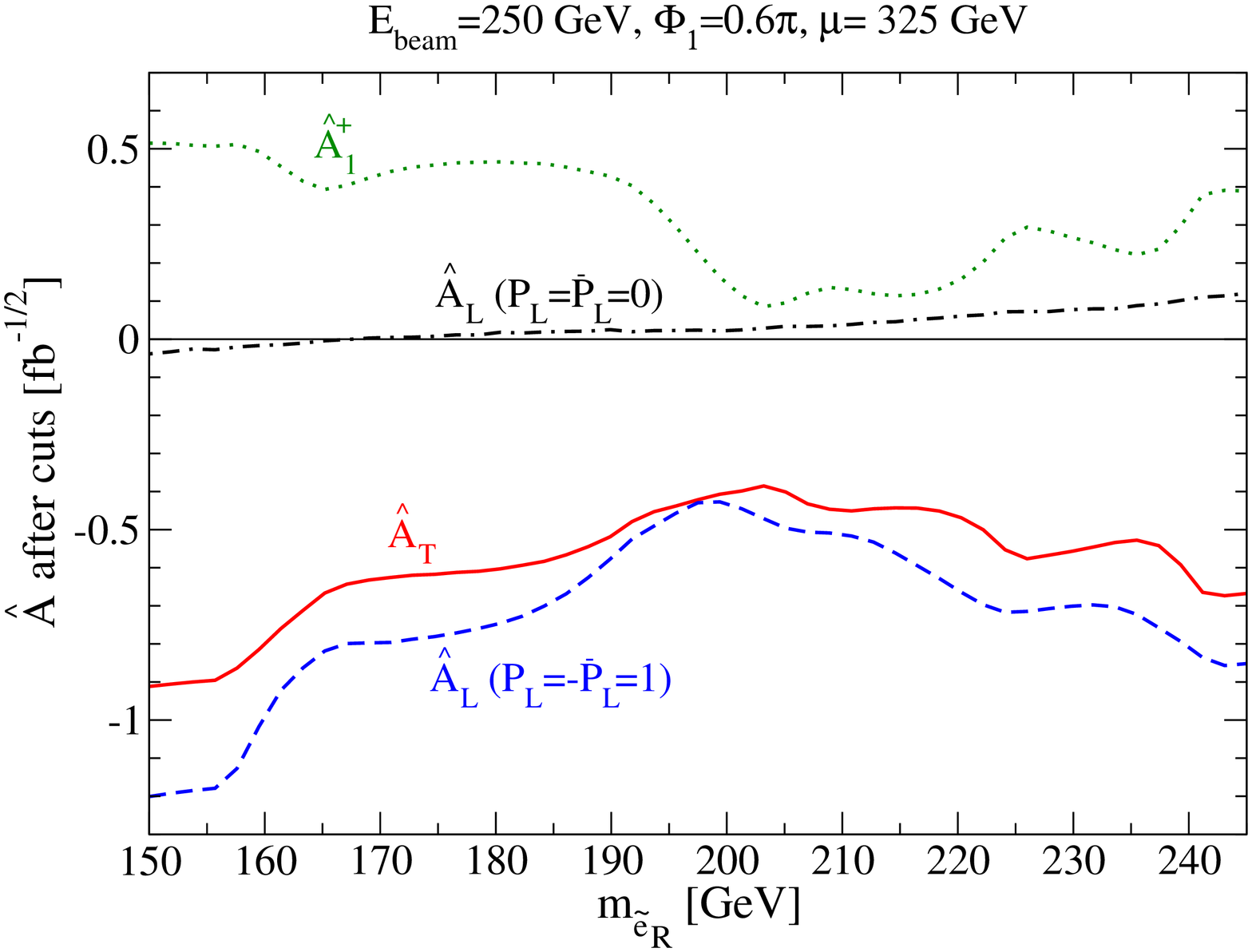}\\
\includegraphics[height=7.8cm,width=7cm,angle=270]{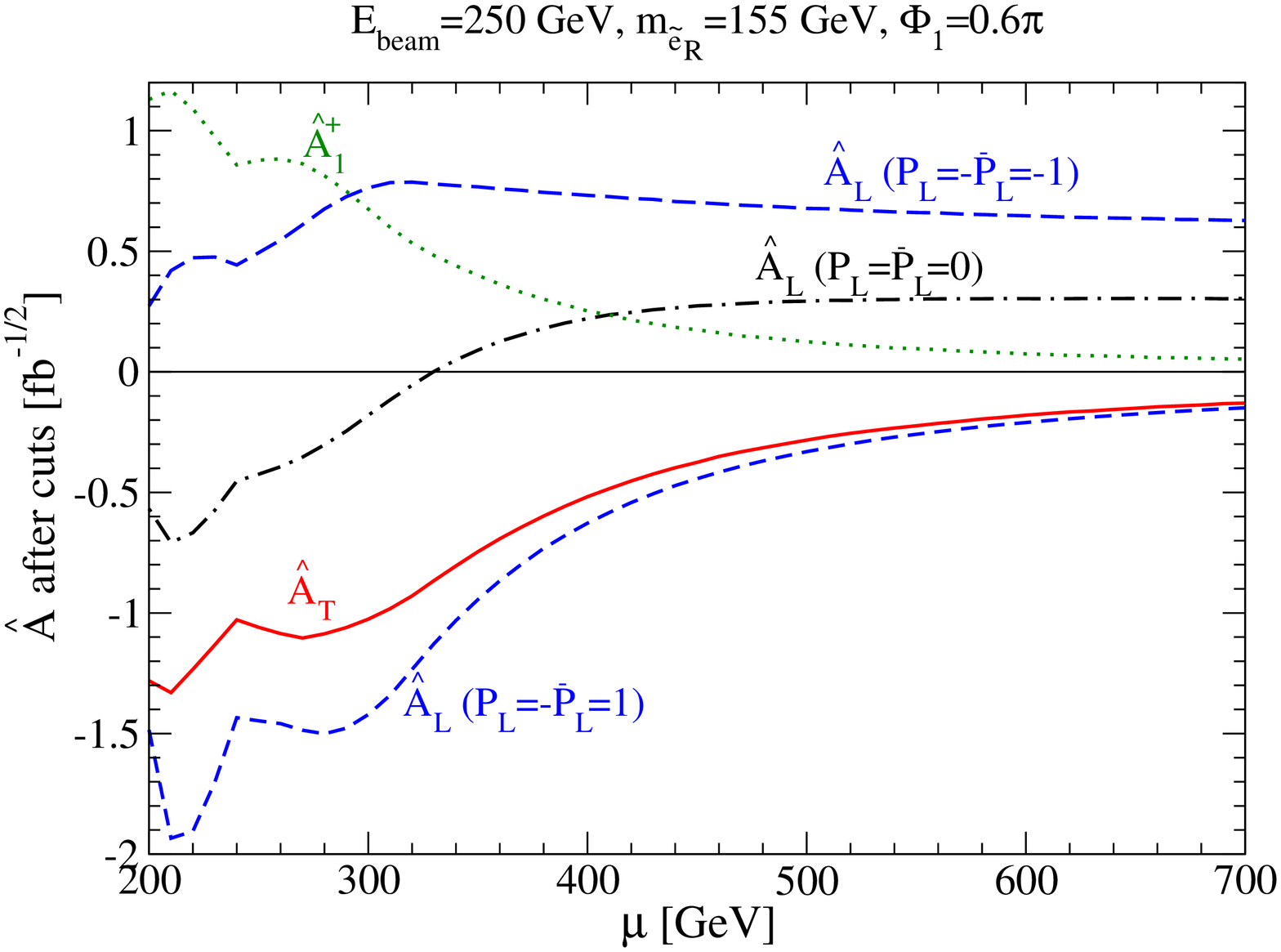} %\hskip 0.5cm
\includegraphics[height=7.8cm,width=7cm,angle=270]{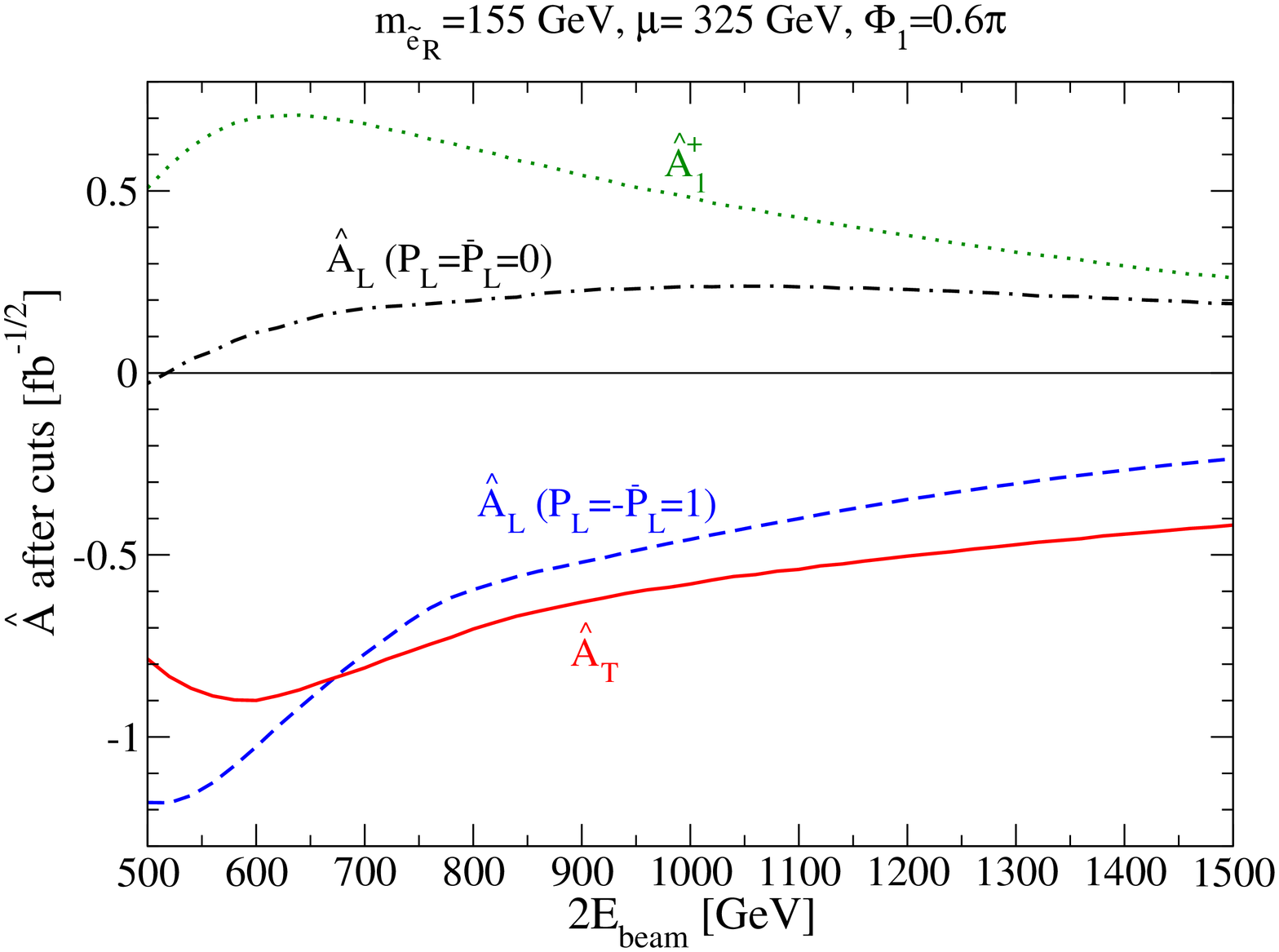}
\end{center}
\vskip -0.5cm
\caption{\it Comparison of the simple transverse decay asymmetry (\ref{f1})
  (green dotted curves), the total ``optimized'' transverse decay asymmetry
  (\ref{ft}) (red solid curves), and the ``optimized'' longitudinal decay
  asymmetry (\ref{fl}), the latter both for transverse (black dot--dashed) and
  for longitudinal (blue dashed) beam polarization. The default values of the
  parameters are as in Fig.~\ref{fig4}, but one parameter is varied in
  each panel.}
\label{fig5}
\end{figure}

Finally, the four figures in Fig.~\ref{fig5} compare the simple asymmetry
$\hat{A}_1^+$ of Eq.~(\ref{f1}), the total optimized transverse decay
asymmetry $\hat{A}_T$, and the optimized longitudinal decay asymmetry
$\hat{A}_L$.  We note that the {\it longitudinal} decay asymmetry
is usually {\it bigger} than our total optimized {\it transverse}
asymmetry. At least for probing the CP-violating phase in the context of
the MSSM (where $\Phi_1$ is the only relevant phase in the convention where
$M_2$ is real), therefore, one does not really seem to gain anything by
transverse beam polarization. The only exception is at large energy
(bottom--right frame); this is due to the extra factor
$m_{\tilde{\chi}_1^0}/\sqrt{s}$ appearing in the expressions for
$\Sigma_{NU}$ in Eq.~(\ref{SBU}), and $\Sigma_{NL}$ in Eq.~(\ref{SAB}),
which determine the size of $\hat A_L$.

The upper right panel shows a quite complicated dependence of the effective
asymmetries on $m_{\tilde e_R}$. For intermediate $\tilde{e}_R$ masses both
final--state leptons in signal events can have similar energies. As a result
one often has four solutions for the event reconstruction.  In this case one
cannot identify the ``primary'' lepton used in Eq.~(\ref{f1}). We have dealt
with this by simply discarding events with four solutions, since averaging
over all four solutions would dilute the asymmetries a lot.  Unfortunately
this reduces the cross section significantly. At the same time
$\tilde{e}_R$ pair events become more similar to our $\tilde{\chi}^0_1
\tilde{\chi}^0_2$ events, since, as we just mentioned, the signal now has
similar distributions for both final $\ell^\pm$ energies.  Hence the cut
against selectron pair production removes more signal events in the present
case. As a result, the complete set of cuts reduces the total cross section by
up to a factor of 5, the worst case being $m_{\tilde e_R} \simeq 195$ GeV.
Note that the different asymmetries are not equally sensitive to these cuts.
The total ``optimized'' transverse decay asymmetry $\hat{A}_T$ is reduced by
at worst a factor of 2, whereas the simple asymmetry $\hat{A}_1^+$ can go down
by a factor of 4. The reason for this is that the cut efficiency depends on
the same production and decay angles that appear in the definitions of our
asymmetries.

The lower left panel includes the longitudinal decay asymmetry
$\hat{A}_L$ for two different choices of longitudinal $e^\pm$ beam
polarization. In both cases we take opposite polarization for the
$e^+$ and $e^-$ beams, since we are dealing with chiral couplings,
see Eq.(\ref{chiral}). Usually taking a right--handed electron beam
is most advantageous, since it maximizes the $\tilde e_R$ exchange
contribution; note that the $\tilde e_R$ coupling to Binos, which is
needed to probe the CP--odd phase $\Phi_1$, is two times larger than
that of $\tilde e_L$. However, for very large $|\mu|$ this choice is
no longer optimal. In this case $\tilde{\chi}^0_2$ becomes more and
more wino--like, i.e., it does not couple to $\tilde{e}_R$. A
right--handed $e^-$ beam means that $\tilde{e}_L$ exchange does not
contribute; the $Z$--exchange contribution also vanishes for
large $|\mu|$. However, taking left--handed electrons one still gets
a sizable contribution from $\tilde{e}_L$ exchange to the cross
section, and also to the asymmetry. In the opposite regime of rather
small $|\mu|$ the asymmetries depend very strongly on this
parameter, since here $\tilde \chi_2^0$ changes from a higgsino-like
to a wino--like state.

As in the previous figures (as well as in Ref.\,\cite{newbartl}) we
took $e^\pm$ beam polarizations $\pm 1$. In the case of longitudinal
beams one can then suppress the $W$ or $\tilde{e}_R$ pair background
(but not both), by appropriate choice of polarization. However, in
practice the beam polarization will be significantly smaller than
this; we therefore left the cuts against both backgrounds in
place. We also note that longitudinal beam polarization can increase
$\hat A_L$ significantly, although the very small size of this
effective asymmetry for our ``default'' parameters and transversely
polarized beams (top left frame) is clearly accidental.

Last but not least, we have checked numerically the effect of varying
the left--handed selectron mass $m_{\tilde{e}_L}$ on the CP--odd
asymmetries. The transverse decay asymmetries, with transversely
polarized beams, are sensitive to the mass; in fact, they get
a bit bigger with smaller mass values. Nevertheless, we have noted
that the longitudinal asymmetry for unpolarized beams becomes much
bigger when the left--handed selectron mass is reduced. For example,
taking parameters as in the top--left frame in Fig.\,\ref{fig5},
except for a reduced $m_{\tilde{e}_L} = 250$ GeV, the maximal value of
$|\hat{A}_T|$ after cuts increases to about 1.2 fb$^{-1/2}$, whereas the
maximum of $|\hat A_L|$ reaches about 2.2 fb$^{-1/2}$. We emphasize
that we do not actually need any beam polarization to probe this
asymmetry, although it can be increased significantly by using
longitudinal polarized beams; for reduced $\tilde{e}_L$ mass, taking
left--handed $e^-$ and right--handed $e^+$ beams is often optimal.
Therefore, reducing the left--handed selectron mass does not affect
the ordering of $A_T$ and $A_L$, i.e. the inequality $A_L> A_T$
(for optimized choice of longitudinal beam polarization.)

%%%%%%%%%%%%%%%%%%%%%%%%%%%%%%%%%%%%%%
\section{Summary and Conclusions}
\label{sec:sec7}
%%%%%%%%%%%%%%%%%%%%%%%%%%%%%%%%%%%%%%

In this paper we studied the production of neutralino pairs at
future linear $e^+e^-$ colliders, with subsequent two--body decays
of the heavier neutralinos. We found that decays of the type $\tilde
\chi_i^0 \to \tilde \chi_j^0 (h,Z)$ are not sensitive to the $\tilde
\chi_i^0$ polarization, unless one can measure the polarization of
the $Z-$boson (or that of the final--state neutralino
$\tilde{\chi}^0_j$). These decays can therefore only be used to probe CP
violation in neutralino {\em production}. Unfortunately the
corresponding CP--odd term suffers from cancelations between $t-$
and $u-$channel diagrams, and is nonzero only in the presence of
higgsino--gaugino mixing. As a result, measuring this asymmetry,
which can be done only with transversely polarized $e^\pm$ beams,
will be very difficult, if not impossible, with the currently
foreseen linear collider performance.

In contrast, $\tilde \chi_i^0$ decays into a slepton plus a lepton allows to
probe the $\tilde \chi_i^0$ polarization state, thereby opening up the
possibility to construct several decay asymmetries.  Moreover, this decay,
followed by subsequent $\tilde \ell \rightarrow \ell \tilde \chi_1^0$ decays,
allows to reconstruct even the simplest neutralino pair events, $\tilde
\chi_2^0 \tilde \chi_1^0$ production with invisible (e.g., stable) $\tilde
\chi_1^0$, with two-- or four--fold ambiguity. Under favorable circumstances
experiments at a collider with (sufficiently strongly) transversely polarized
beams should then be able to determine non--vanishing asymmetries with high
statistical significance.  However, even in this case a different asymmetry,
which does not depend on transverse beam polarization (but can be maximized
using longitudinal beam polarization), is generally larger in size than even
the best of the transverse decay asymmetries we studied. We saw in
Fig.~\ref{fig5} that this is true both for gaugino-- and higgsino--like
$\tilde \chi_2^0$. It also remains true when we vary the ratio $|M_1|/M_2$, in
particular for $|M_1| > M_2$. However, if $|M_1| \gg M_2, \, |\mu|$, or if
both produced neutralinos are higgsino--like, all CP--odd asymmetries become
small. Recall that in the MSSM all these asymmetries essentially result from
a single (potentially large) phase, associated with the U(1) gaugino mass (in
the convention where the SU(2) gaugino mass is real and positive).

We therefore conclude that,   {\it at least} in the context of neutralino
production in the MSSM, transverse beam polarization is not particularly
useful in probing explicit CP violation. Once the relevant masses have been
determined, the most sensitive probe of the relevant CP--odd phases remains
the total cross section \cite{Choi:2004rf}, although it is a CP--even observable.
If this measurement indicates that some phase
differs from 0 or $\pi$, one needs to see explicit CP violation,  in order
to convince oneself that the variation of the cross section is indeed due to a phase,
rather than due to some extension of the MSSM. However, as noted above, this
can be most easily accomplished by using longitudinal, rather than transverse,
beam polarization.

The situation might be different in extensions of the MSSM, however. Whenever
the quartic charges $Q_6$ and $Q_6^\prime$ defined in Sec.~2.2
%that can be probed with transversely polarized $e^\pm$ beams
contain (combinations of) phases that are independent of those
%that can be probed with longitudinal beam polarization,
in $Q_4$ and $Q_4^\prime$, the option of transverse beam
polarization might be very useful for determining these phases.
In the NMSSM, for example,
the neutralino mass matrix contains additional CP--odd phases associated with
the singlino sector, which can be large. A dedicated analysis along
the lines presented in this paper would be required to decide whether
transverse beam polarization could be helpful in disentangling this more
complicated neutralino sector.

\subsection*{Acknowledgments}

We thank Saurabh Rindani for discussions that triggered this investigation,
and Peter Zerwas for discussions and suggestions. The work of JS was
supported by the Korea Research Foundation Grant (KRF-2005-070-C00030). The
work of SYC was supported partially by the Korea Research Foundation Grant
(KRF--2004--041--C00081) and by KOSEF through CHEP at Kyungpook National
University. MD thanks the Center for Theoretical Physics at Seoul National
University, as well as the particle theory group at the University of Hawaii
at Manoa, for hospitality.


\begin{thebibliography}{99}

\bibitem{book} M. Drees, R.M. Godbole and P. Roy, {\it Theory and
    Phenomenology of Sparticles}, World Scientific (Singapore, 2004); D.J.H.
  Chung, L.L. Everett, G.L. Kane, S.F. King, J. Lykken and L.T. Wang, Phys.
  Rep. {\bf 407}, 1 (2005).

\bibitem{edm} J.R. Ellis, S. Ferrara and D.V. Nanopoulos, Phys. Lett.
   B {\bf 114}, 231 (1982); F. del Aguila, M.B. Gavela, J.A. Grifols and
   A. Mendez, Phys. Lett. B {\bf 126}, 71 (1983), Erratum-ibid. B {\bf 129},
   473 (1983).

\bibitem{edm1} T. Ibrahim and P. Nath, Phys. Lett. B {\bf 418}, 98 (1998);
   Phys. Rev. D {\bf 57}, 478 (1998); D {\bf 58}, 019901 (1998) (E);
   {\it ibid}, 111301 (1998); {\it ibid.} D {\bf 61}, 095008 (2000),
   hep--ph/9907555; M. Brhlik, G.J. Good and G.L. Kane, Phys. Rev. D {\bf 59},
   115004 (1999), hep--ph/9810457; M. Brhlik, L.L. Everett, G.L. Kane and
   J.D. Lykken, Phys. Rev. Lett. {\bf 83}, 2124 (1999), hep--ph/9905215.

\bibitem{Choi:2004rf} S.Y. Choi, M. Drees and B. Gaissmaier, Phys. Rev.
   D {\bf 70}, 014010 (2004), hep--ph/0403054.

\bibitem{others} S.T. Petcov, Phys. Lett. B {\bf 178}, 57 (1986);
   Y. Kizukuri and N. Oshimo, Phys. Lett. B {\bf 249} (1990) 449;
   V. Barger, T. Han, T. Li and T. Plehn, Phys. Lett. B {\bf 475},
   342 (2000), hep--ph/9907425; V.D. Barger, T. Falk, T. Han, J. Jiang,
   T. Li and T. Plehn, Phys. Rev. D {\bf 64}, 056007 (2001), hep--ph/0101106;
   J. Kalinowski, Acta Phys. Polon. B {\bf 34}, 3441 (2003), hep--ph/0306272;
   A. Bartl, H. Fraas, O. Kittel and W. Majerotto, Phys. Rev. D {\bf 69},
   035007 (2004), hep--ph/0308141, and Eur. Phys. J. C {\bf 36}, 233 (2004),
   hep--ph/0402016; A. Bartl, H. Fraas, S. Hesselbach, K. Hohenwarter--Sodek
   and G. Moortgat--Pick, JHEP {\bf 0408}, 038 (2004), hep--ph/0406190;
   S.Y. Choi, Phys. Rev. D {\bf 69}, 096003 (2004), hep--ph/0308060.

\bibitem{CSS} S.Y. Choi, H.S. Song and W.Y. Song, Phys. Rev. D {\bf 61}, 075004
   (2000), hep--ph/9907474.

\bibitem{cdgs} A. Bartl, T. Kernreiter and O. Kittel, Phys. Lett. B {\bf 578},
  341 (2004), hep--ph/0309340; S.Y. Choi, M. Drees, B. Gaissmaier and J. Song,
   Phys. Rev. {\bf D69}, 035008 (2004), hep--ph/0310284.

\bibitem{newbartl} A. Bartl, H. Fraas, S. Hesselbach, K. Hohenwarter-Sodek,
   T. Kernreiter and G. Moortgat-Pick, hep--ph/0510029.

\bibitem{oldino} J.R. Ellis, J.M. Fr\`ere, J.S. Hagelin, G.L. Kane and
   S.T. Petcov, Phys. Lett. B {\bf 132}, 436 (1983); V. Barger, R.W. Robinett,
   W.Y. Keung and R.J.N. Phillips, Phys. Lett. B {\bf 131}, 372 (1983);
   A. Bartl, H. Fraas and W. Majerotto, Nucl. Phys. B {\bf 278}, 1 (1986),
   and Z. Phys. C {\bf 30}, 441 (1986); G. Moortgat--Pick and H. Fraas,
   Phys. Rev. D {\bf 59}, 015016 (1999), hep--ph/9708481];
   G. Moortgat--Pick, H. Fraas, A. Bartl and W. Majerotto, Eur. Phys. J. C
   {\bf 9}, 521 (1999), Erratum-ibid. C {\bf 9}, 549 (1999), hep--ph/9903220;
   G. Moortgat-Pick, A. Bartl, H. Fraas and W. Majerotto,
   Eur. Phys. J. C {\bf 18}, 379 (2000), hep--ph/0007222.

\bibitem{rec} T. Tsukamoto, K. Fujii, H. Murayama, M. Yamaguchi and Y. Okada,
   Phys. Rev. D {\bf 51}, 3153 (1995); J.L. Feng, M.E. Peskin, H. Murayama and
   X. Tata, Phys. Rev. D {\bf 52}, 1418 (1995), hep--ph/9502260:
   H. Baer, R. Munroe and X. Tata, Phys. Rev. D {\bf 54}, 6735 (1996),
   Erratum-ibid. D {\bf 56}, 4424 (1997), hep--ph/9606325;
   J.L. Kneur and G. Moultaka, Phys. Rev. D {\bf 59}, 015005 (1999),
   hep--ph/9807336, and Phys. Rev. D {\bf 61}, 095003 (2000), hep--ph/9907360;
   G.A. Blair, W. Porod and P.M. Zerwas, Phys. Rev. {\bf D63}, 017703 (2001),
   hep--ph/0007107.

 \bibitem{Choi:2001ww} S.Y. Choi, J. Kalinowski, G. Moortgat-Pick and P.M.
   Zerwas, Eur. Phys. J. C {\bf 22}, 563 (2001); {\it ibid.} C {\bf 23}, 769
   (2002).

\bibitem{inoloop} D. Pierce and A. Papadopoulos, Phys. Rev. D {\bf 50}, 565
   (1994), hep--ph/9312248, and Nucl. Phys. B {\bf 430}, 278 (1994),
   hep--ph/9403240; A.B. Lahanas, K. Tamvakis and N.D. Tracas, Phys. Lett.
   B {\bf 324}, 387 (1994), hep--ph/9312251; H. Eberl, M. Kincel, W. Majerotto
   and Y. Yamada, Phys. Rev. D {\bf 64}, 115013 (2001), hep--ph/0104109;
   T. Fritzsche and W. Hollik, Eur. Phys. J. C {\bf 24}, 619 (2002),
   hep--ph/0203159; W. Oller, H. Eberl, W. Majerotto and C. Weber,
   Eur. Phys. J. C {\bf 29}, 563 (2003), hep--ph/0304006;
   W. Oller, H. Eberl and W. Majerotto, Phys. Lett. B {\bf 590}, 273 (2004),
   hep--ph/0402134.

 \bibitem{Hagiwara:1985yu} K. Hagiwara and D. Zeppenfeld, Nucl. Phys. B {\bf
     274}, 1 (1986); G.A. Ladinsky, Phys. Rev. D {\bf 46}, 2922 (1992).

\bibitem{Choi:2003fs} J.F. Gunion and H.E. Haber, Phys. Rev. D {\bf 37}, 2515
   (1988); S.Y. Choi and Y.G. Kim, Phys. Rev. D {\bf 69}, 015011 (2004).

\bibitem{abdel} A. Djouadi, M. Drees and J.-L. Kneur, JHEP {\bf 0108}, 055
   (2001), hep--ph/0107316.

\bibitem{optimal} D. Atwood and A. Soni, Phys. Rev. D {\bf 45}, 2405 (1992);
   M. Diehl and O. Nachtmann, Z. Phys. C {\bf 62}, 397 (1994); M. Davier,
   L. Duflot, F. Le Dieberder and A. Roug\'{e}, Phys. Lett. B {\bf 306}, 411
   (1993); J.F. Gunion, B. Grzadkowski and X.-G. He, Phys. Rev. Lett.
   {\bf 77}, 5172 (1996).

\bibitem{bartl_char} A. Bartl, K. Hohenwarter--Sodek, T. Kernreiter and
   H. Rud, Eur. Phys. J. C {\bf 36}, 515 (2004), hep--ph/0403265.

\end{thebibliography}
\end{document}